\documentclass[aps,prd,
preprint,
nofootinbib,groupedaddress]{revtex4-2}

\usepackage{amsmath}
\usepackage{amsfonts}
\usepackage{graphicx}
\usepackage{hyperref}
\usepackage{caption}
\usepackage{slashed}
\usepackage[mathscr]{euscript}

\usepackage{lineno}
\usepackage{amssymb}
\usepackage{amsmath}
\usepackage{graphicx}
\usepackage{dcolumn}
\usepackage{bm}
\usepackage{epsfig}
\usepackage{amssymb} 
\usepackage{datetime}
\usepackage{wasysym}
\usepackage{slashed} 
\usepackage[mathscr]{euscript}
\usepackage{ulem}

\usepackage{microtype}

\newcommand{\bal}{\begin{align}}
\newcommand{\eal}{\end{align}}
\newcommand{\beq}{\begin{eqnarray}}
\newcommand{\eeq}{\end{eqnarray}}
\newcommand{\nneeq}{\nonumber \end{eqnarray}}

\newcommand{\nn}{\nonumber \\}
\newcommand{\es}{& = &}
\newcommand{\rs}{\, = \,}
\newcommand{\ps}{& + &}
\newcommand{\ms}{& - &}
\newcommand{\ts}{& \times &}
\newcommand{\nt}{\nn \ts}
\newcommand{\np}{\nn \ps}
\newcommand{\nm}{\nn \ms}

\newcommand{\cM}{ {\mathcal M} }
\newcommand{\cH}{ {\mathcal H} }
\newcommand{\cE}{ {\mathcal E} }
\newcommand{\cF}{ {\mathcal F} }
\newcommand{\cG}{ {\mathcal G} }

\newcommand{\cU}{ {\mathcal U} }

\newcommand{\cP}{ {\mathcal P} }

\newcommand{\tdelta}{\tilde\delta}
\newcommand{\pd}{ {\partial} }
\newcommand{\ket}[1]{ {\left|{#1}\right\rangle} }
\newcommand{\bra}[1]{ {\left\langle{#1}\right|} }

\newcommand{\bmat}{\left[\begin{array}}
\newcommand{\emat}{\end{array}\right]}

\makeatletter
\renewcommand{\beq}{%
 \begingroup
 \eqnarray%
 \@ifstar{\nonumber}{}%
}
\renewcommand{\eeq}{\endeqnarray\endgroup}
\makeatother

\begin{document}


\title{Dynamics of heavy quarks in the Fock space}


\author{Kamil Serafin}
\affiliation{Institute of Modern Physics, Chinese Academy of Sciences, Lanzhou 730000, China}
\affiliation{CAS Key Laboratory of High Precision Nuclear Spectroscopy, Institute of Modern Physics, Chinese Academy of Sciences, Lanzhou 730000, China}
\author{Mar\'ia G\'omez-Rocha}
\affiliation{Departamento de Física Atómica, Molecular y Nuclear and Instituto Carlos I de Física Teórica y Computacional, Universidad de Granada, Granada, Spain}
\author{Jai More}
\affiliation{Department of Physics, Indian Institute of Technology Bombay, Powai, Mumbai, 400076, India}
\author{S. D. Głazek}
\affiliation{Institute of Theoretical Physics, Faculty of Physics, University of Warsaw}


\date{\today}

\begin{abstract}
This paper concerns a method of describing hadrons that starts with the canonical front form Hamiltonian of QCD. The method is developed in the relatively simple context of QCD with only heavy quarks. We regulate its canonical Hamiltonian by introducing a vanishingly small gluon mass $m_g$. For positive $m_g$, the small-$x$ gluon divergences become ultraviolet and hence they are renormalized in the same way the ultraviolet transverse divergences are. This is done using the renormalization group procedure for effective particles. Up to the second order of expansion of the renormalized Hamiltonian in powers of the quark-gluon coupling constant $g$, only the quark mass-squared and gluon-exchange divergences require counter terms. In these circumstances, we calculate an effective potential between quarks in heavy quarkonia in an elementary way, replacing all the quarkonium-state components with gluons 
of mass $m_g$ by only one component with just one gluon that is assigned a mass $m_G$, comparable to or exceeding the scale of typical relative momenta of bound quarks. In the limit of $m_g \to 0$ and large $m_G$ two results are obtained. (1) While the color-singlet quarkonium mass eigenvalue stays finite and physically reasonable in that limit, the eigenvalues for single quarks and octet quarkonia are infinite. (2) The effective quark-antiquark potential is quadratic as a function of the distance and spherically symmetric for typical separations between quarks but becomes logarithmic and no longer spherically symmetric for large separations. Our conclusion indicates how to systematically improve upon the approximations made in this paper. 
\end{abstract}


\maketitle

\section{Introduction}

Description of heavy-quark bound states in terms of their 
virtual Fock-space components is meant to be achievable 
through solving the Hamiltonian eigenvalue problem in QCD, 
which in the first approximation is limited to only involve 
quarks $b$ and $c$. However, the canonical Hamiltonian of 
even so severely limited theory poses conceptual and computational 
problems. To begin with, the Hamiltonian needs regularization. 
The formal momentum cutoffs one imposes on the virtual Fock 
states of heavy quarks are much greater than the quark 
masses. Therefore, from the regularization point of view, 
the heavy quarks do not differ much from the light ones -- 
their masses are negligible in comparison with the cutoffs. 
Further, the canonical gluon mass is zero, which is infinitely
small in comparison to any non-zero quark mass. The key
distinction  between the heavy and light quarks is provided
by the ratio of their masses to the parameter $\Lambda_{\rm QCD}$.
The latter results from dimensional transmutation~\cite{Coleman:1973jx,Gross:1973id,Politzer:1973fx}. 
However, such a parameter cannot be introduced in a precise
way without a renormalization group procedure for Hamiltonians. 
One faces the difficulty that Hamiltonians in quantum field 
theory result from integrating Hamiltonian densities over a
three-dimensional space. The three-dimensional Hamiltonian 
density is a different concept from the four-dimensional 
Lagrangian density and the associated concepts of action, 
path integration and diagrammatic techniques of perturbation 
theory.

In this paper, we address the issue of dynamics of heavy quarks 
using the renormalization technique called the renormalization 
group procedure for effective particles (RGPEP). The RGPEP is 
designed for Hamiltonians and it applies to the front form (FF) 
of dynamics~\cite{Dirac1949,Brodsky:1997de}. One is motivated to 
use the FF instead of the more familiar instant form (IF) because 
of the desire to describe the quarkonia observed in motion as 
well as the quarkonia observed at rest and to include other 
moving particles with which the heavy quarks interact. The key 
feature of the FF of Hamiltonian dynamics is that the required 
boosts are kinematic; their generators do not involve interactions. 
This is not the case in the IF, where motion is associated with 
a dynamical change in the virtual Fock-space decomposition. 

The RGPEP has been used for the purpose of describing heavy quarks before~\cite{QbarQ,Serafin_Baryons,Serafin:2019vuk}, 
including the effective potentials derived using the RGPEP that were used in other approaches for description of heavy tetraquarks ~\cite{Kuang:2022vdy}. The new element utilized in this article is the small gluon mass parameter $m_g$ that regulates the singularities caused by gluons carrying small longitudinal momenta~\cite{Glazek:2018kvx,Galvez-Viruet:2023bid} or, in the parton-model language, those that carry a small $x$. The arbitrarily small parameter $m_g$, in combination with the running scale parameter, denoted by $t$, provides a lower bound on $x$. This bound is absent in the free part of the Hamiltonian. It only appears in the interaction terms. The reason is that the RGPEP does not integrate out large FF energies. Instead it integrates out large {\it changes} of the FF energies due to interactions. The free part of the Hamiltonian does not change the FF energy. In other words, instead of Wilsonian integrating out of large energies and hence limiting the range of momenta of field quanta in the Fock-space basis, we only integrate out the interactions that cause large changes of the FF energy. 
In this respect, our Hamiltonian approach differs from the extended literature on heavy-quark bound states, including the outstanding reviews~\cite{Brambilla,Pineda}.
Regarding boost invariance in Minkowski approaches, we wish to mention~\cite{Hilger:2014nma,FischerKubrak,Leitao:2016bqq,Bashir:1998wd,Gomez-Rocha:2016cji} and references therein, from which our Hamiltonian approach also differs in this respect.

As a consequence of the regularization used in this article, one circumvents the vacuum problem in the theory. Instead of involving the ground state in the dynamics and arguing that it somehow expels the colored states out from the spectrum, we find that the renormalized Hamiltonian is capable of producing infinite eigenvalues for colored states in the limit $m_g \to 0$. This result only follows under the assumption that an exchange of an effective gluon between effective quarks is blocked by the non-Abelian interactions when the RGPEP running scale parameter $t$ is increased to the value that characterizes the formation of the quark bound states. 

The simplest model of the blocking of effective gluons from being exchanged between effective quarks is defined by giving the gluons a large mass $m_G \gg \Lambda_{\rm QCD} \gg m_g$, where $\Lambda_{\rm QCD}$ is defined in the RGPEP scheme.  One does not need to specify how $m_G$ depends on $t$. It suffices to assume that it is larger than the momentum transfers between quarks involved in formation of bound states. That way the Fock-space dynamics, which {\it a priori} involves unlimited numbers of effective gluons, is drastically simplified because emission and absorption of heavy gluons is suppressed. In our calculation, we limit the number of heavy effective gluons involved in the dynamics to one.
Since our FF Hamiltonian approach is only developed in gauge $A^+=0$, we ought to mention that an effective gluon mass has also been found useful in calculations using the Dyson-Schwinger equations in Landau gauge, see Refs.~\cite{Aguilar:2021uwa,Aguilar:2015bud,Binosi:2017rwj,Aguilar:2019kxz}.

The results we report follow from the renormalized Hamiltonian that is computed in the limit $m_g \to 0$ using expansion in a series of powers of the coupling constant only up to second order. Although nowadays it may be not surprising that such low-order computations can point toward some mechanism of quark binding~\cite{Perry:1994mv,BrisudovaPerryWilson,QQ1}, the mechanism our calculation points to is surprisingly simple: the quark self-interaction tends to infinity as $|\log(t m_g^2)|$, but this logarithmic growth is canceled in colorless quarkonia by the effective interaction term computed using the RGPEP. 

Before the limit $m_g \to 0$ is taken, a finite value of $m_g$ converts the small-$x$ divergences into the large FF energy divergences. It happens because the gluon minus momentum is
\beq
p_g^-
\es
\frac{ m_g^2 + p_g^{\perp 2} }{ p_g^+ } \ .
\eeq
It becomes infinite when $p_g^+ \to 0$ no matter how small $p_g^\perp$ is. If $m_g$ is zero, $p_g^\perp \sim \sqrt{x}$ or smaller would lead to finite or even vanishing $p_g^-$. However, for $m_g > 0$, the divergences due to small $x = 
p_g^+/P^+$, where $P^+$ is a momentum of a system under consideration, can be treated on an equal footing with the transverse UV divergences associated with $p_g^\perp \to \infty$, 
using the RGPEP.

It should be pointed out that the finite and phenomenologically reasonable eigenvalues we obtain for white quarkonia depend on the ratio of the quarkonium longitudinal momentum as a whole, $P^+$, to the running renormalization group scale-parameter $t$. This is so due to the approximations we are forced to make at this stage of developing the theory. Namely: instead of solving the RGPEP equation for scale-dependent, effective Hamiltonians $H_t$ exactly, we use expansion in powers of $g$ only up to $g^2$; 
we introduce the hypothetical mass $m_G$ for effective gluons, while we do not know yet how the effective gluon mass actually evolves with $t$; we limit the eigenvalue problem for $H_t$ to a subspace spanned by two effective components, one with a quark and an antiquark and another one with a quark, an antiquark and a gluon; we approximate the dynamical contribution of the three-particle component to the eigenvalue equation for 
the two-particle component keeping only terms on the order of $g^2$; we adjust the value of $t$ to bring the resulting mass eigenvalues close to data, because our calculations are not exact so that the resulting observables depend on $t$, and we extrapolate the coupling constant $g$ to the value $g_t$ that is in the ballpark of expectations based on the perturbative evolution of $g_t$, but we cannot estimate the magnitude of error caused by such extrapolation in calculations limited to second order.

The paper is organized in the following way. Section~\ref{sec:Hgluonmass} 
introduces the canonical FF Hamiltonian of QCD to which we add the gluon mass term with $m_g$. In Section~\ref{sec:reno} we apply the RGPEP to compute and then supplement with $m_G$ the effective Hamiltonian for a sizable $t$, keeping terms on the order of 1, $g$ and $g^2$. The quarkonium eigenvalue problem for the resulting Hamiltonian $H_t$ is examined in Section~\ref{sec:quarkonium}. The effective potential we obtain in the nonrelativistic limit of the eigenvalue problem is described in Section~\ref{sec:nonrel}. {\ Appendix~\ref{ap:SecondOrder} contains details of derivation of the effective Hamiltonian up to the second order, while Appendix~\ref{sec:gluonExchCounterterm} shows that the gluon exchange counterterm ensures cancellation of small-x divergences due to exchange of a gluon.} We often abbreviate the quark quantum number subscripts, momentum, isospin or flavor, spin and color in just one subscript. For example, instead of $p_1$, $i_1$ or $f_1$, $\sigma_1$ and $c_1$, we write only 1.

\section{QCD with gluon mass $m_g$}
\label{sec:Hgluonmass}

The canonical FF Hamiltonian density for QCD is obtained
from its classical Lagrangian through the well-known 
quantization procedure in gauge $A^+=0$, {\it e.g.} 
see~\cite{Brodsky:1997de}. We limit the theory to quarks 
$c$ and $b$ and supply the Hamiltonian density with a 
gluon mass term,
\beq
\cH
\es
  \cH_\mathrm{QCD}
+ \frac{1}{2} m_g^2 A^{\perp\,a} A^{\perp\,a}
\ .
\eeq
The mass $m_g$ can be considered extremely small,
so that its presence in a regulated quantum theory is 
not noticeable at the level of classical gauge symmetry
and could not be detected by experiment (current upper
limit on the gluon mass is on the order of a few MeV/$c^2$
\cite{Workman:2022ynf}). After integration over $x^-$ and 
$x^\perp$ and normal-ordering, one obtains the free 
Hamiltonian term for gluons in the form
\beq
H_{A^2}
\es
\int_3 p_3^- a_3^\dag a_3
\ ,
\label{eq:canHfGluon}
\eeq
where $a_3$ is the gluon annihilation operator labeled by
quantum numbers $\sigma_3$, $c_3$, $p_3^+$, and $p_3^\perp$ 
for polarization, color, longitudinal momentum and transverse 
momentum, respectively, collectively denoted by $3$.
\beq
p_3^-
\es
\frac{ m_g^2 + p_3^{\perp 2} }{ p_3^+ }
\ ,
\label{eq:p_3^-}
\eeq
and
\beq
\int_3
\es
\sum_{\sigma_3, c_3} \int\frac{dp_3^+ d^2p_3^\perp}{16\pi^3 p_3^+}
\ .
\eeq
We use label $3$ for gluons because we choose label $1$ for quarks
and label $2$ for antiquarks. For example, labels $1'$ and $\tilde 1$
will refer to quarks. Also, $\int_1 \int_2 = \int_{12}$ and phrase
``pair 12'' refers to a quark 1 and antiquark 2. The invariant mass 
squared of a quark-gluon pair $13$ is $\cM_{13}^2 = (p_1^+ + p_3^+)
(p_1^- + p_3^-) - (p_1^\perp + p_3^\perp)^2$ with the minus components 
being eigenvalues of the free part of the canonical Hamiltonian.

Nonzero $m_g$ implies that whenever $p_3^+$ approaches zero, $p_3^-$ approaches infinity. In contrast, if $m_g = 0$, one can simultaneously 
set $p_3^+ \to 0$, and $p_3^\perp \to 0$ in such a way that $p_3^-$
stays constant or vanishes. Therefore, for $m_g > 0$ all gluon modes 
with vanishing longitudinal momenta are FF high-energy (large $p_3^-$) 
modes, while for $m_g = 0$ such modes can also be FF small-energy 
modes. Using $m_g > 0$ one can simultaneously regulate transverse 
and small-$x$ singularities by a cutoff on the invariant mass. 
Renormalization due to both singularities is discussed in Sec.~\ref{sec:reno}. Once the Hamiltonian is renormalized so that the cutoff dependence 
of its matrix elements between states of effective particles with 
finite momenta is removed, the resulting theory depends on $m_g$. 
We discuss its limit when $m_g \to 0$ in Sec.~\ref{sec:quarkonium} 
on the example of quarkonium eigenvalue problem.

\section{Renormalized Hamiltonian\label{sec:reno}}

To obtain the renormalized Hamiltonian we use the renormalization
group procedure for effective particles (RGPEP). The procedure 
involves regularization, calculation of an effective Hamiltonian
and determination of counterterms, if necessary.

Gluon operators $a_3$ (and $a_3^\dag$), introduced in 
Eq.~(\ref{eq:canHfGluon}), annihilate (and create) gluons
that can be characterized as bare or point-like. Effective 
gluons are created by $a_{t \, 3}^\dag$ and annihilated 
by $a_{t \, 3}$, where $t \ge 0$ is the RGPEP parameter 
related to the size of effective particles. The canonical, 
point-like gluons correspond to $t = 0$. The same relationships
hold for quark operators. So, the effective quarks and antiquarks 
are created by $b_{t \, 1}^\dag$ and $d_{t \, 2}^\dag$, respectively. 
The effective operators are related to the bare operators by means 
of a unitary transformation $\cU_t$,
\beq
b_{t \, 1} \rs \cU_t b_1 \cU_t^\dag
\, ,
\quad
d_{t \, 2} \rs \cU_t d_2 \cU_t^\dag
\, ,
\quad
a_{t \, 3} \rs \cU_t a_3 \cU_t^\dag
\, .
\eeq
Dimension of the parameter $t$ is the front form energy 
to power minus two. In accordance with the dimensional 
analysis of Ref.~\cite{Wilson:1994fk}, we introduce two 
scale parameters, longitudinal momentum scale $\cP$, 
and transverse momentum scale $\lambda$,
\beq
t
\es
\frac{\cP^2}{\lambda^4}
\ .
\label{eq:tlambda}
\eeq

The effective Hamiltonians, denoted by $H_t$, are linear 
combinations of products of creation and annihilation
operators for effective particles. The coefficients in 
front of those products are functions of $t$ as well 
as all quantum numbers describing each particle operator 
involved. The effective Hamiltonians describe the same 
theory, thus $H_t = H_0$. Additionally, we define
\beq
\cH_t
\es
\cU_t^\dag  H_{ t}  \, \cU_t
\ .
\label{eq:defcHt}
\eeq
Whereas $H_t$ is a linear combination of products of effective
operators, $\cH_t$ is the same linear combination (with the same
coefficients) of products of bare particle operators. 
We are able now to implicitly define $\cU_t$ by demanding that $\cH_t$
is the solution of the following differential equation, 
\beq
\frac{d\cH_t}{dt}
\es
\left[ \left[ \cH_f, \cH_t \right], \, \cH_t \right] ,
\label{eq:RGPEPeq1}
\eeq
where $\cH_f$ is the free part of $\cH_t$, i.e., the part that
is obtained by setting the coupling constant $g$ to 0. The relation
between $\cU_t$ and $\cH_t$ can be recovered from Eqs.~(\ref{eq:defcHt})
and (\ref{eq:RGPEPeq1}) remembering that $dH_t/dt = 0$. Effective 
particles essentially define a $t$-dependent basis in the space of
states. Equation~(\ref{eq:RGPEPeq1}) is simpler than the corresponding equation in Ref.~\cite{Glazek:2012qj} and it leads to simplified 
computations in the cases we consider. We adopt the simplification 
because it readily yields the attractive results that are described 
in this paper.

The operator $H_{t>0}$ that solves Eq.~(\ref{eq:RGPEPeq1}) order-by-order 
in powers of the coupling constant $g$ is "narrow" in terms of creation and annihilation 
operators for effective particles~\cite{Glazek:2012qj} in the sense similar to the 
narrowness of solutions to the Wegner equation for Hamiltonian matrices \cite{Wegner1, Wegner2}. It means that the matrix elements of $H_t$ between effective particle states 
with vastly different FF energies are negligibly small. More precisely, if the difference 
of FF energies considerably exceeds $t^{-1/2} = \lambda^2/\cP$, then the matrix element 
is exponentially suppressed and equivalent to zero in our calculations. Therefore, the 
larger $t$ the narrower the Hamiltonian. Consequently, one can apply to $H_t$ the 
principles of the similarity renormalization group procedure~\cite{Glazek:1993rc}. 

\subsection{Formulas for the Hamiltonian}

Below we present the relevant part of the renormalized Hamiltonian obtained as
an approximate solution of Eq.~(\ref{eq:RGPEPeq1}) with
the regularized canonical Hamiltonian of heavy-flavor
QCD as the initial condition. The solution is obtained
using the power expansion in the coupling constant $g_t$
for some finite $t$, although up to the second order
$g_t = g$, where $g$ is the coupling constant of the
canonical Hamiltonian. Nevertheless, we use the notation $g_t$ 
to indicate that the coupling constant will evolve with $t$ in higher order calculations than the ones described in this paper. 

We do not present the formulas for the pure canonical 
Hamiltonian. They can be recovered by putting $t = t_r = 0$ 
and omitting the counter terms.

Interaction vertices are regularized.
\begin{figure}
\includegraphics[scale=0.6]{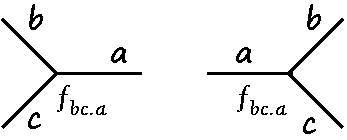}
\caption{First-order interaction vertices. The left vertex stands for annihilation of particle $a$ and creation of particles $b$, $c$ and 
the right vertex is for annihilation of particles $b$, $c$ and creation of particle $a$. The regularization factors $f_{bc.a, t_r}$ in these vertices
are the same. \label{fig:threeleg}}
\end{figure}
For a three-leg vertex in which particle $a$ is annihilated 
and particles $b$ and $c$ are created, and for its hermitian 
conjugate, see Fig.~\ref{fig:threeleg}, the regulating factor 
is set to
\beq
f_{bc.a, t_r}
\es
\exp\left[ - t_r \left( p_b^- + p_c^- - p_a^- \right)^2  \right]
\\
\es
\exp\left[ - t_r \left( \frac{\cM_{bc}^2 - m_a^2}{p_a^+} \right)^2  \right] ,
\label{eq:regulator}
\eeq
where $\cM_{bc}^2$ is the invariant mass of the $bc$ pair,
$m_a$ is the mass of particle $a$, and $t_r$ is a positive
regulating parameter. In the renormalized Hamiltonian, after 
cancellations of divergences are guaranteed, one can set
$t_r$ to zero. The Hamiltonian includes also the instantaneous 
interaction terms that graphically have four legs. Consider 
a term that changes particles 1' and 2' to 1 and 2, see 
Fig.~\ref{fig:exchange}. It is interpreted as composed of 
two three-leg vertices labeled by 1, 1', 3 and 2, 2', 3, 
respectively, and joined by the common leg 3. The corresponding 
regularization factor is set to 
\beq
r_{12.1'2'}
\es
  \theta\left( p_{1'}^+ - p_1^+ \right)
  f_{13.1', \, t_r} \, f_{2'3.2, \, t_r}
\np
 \theta\left( p_1^+ - p_{1'}^+ \right)
  f_{1'3.1, \, t_r} \, f_{23.2', \, t_r}
\ .
\eeq
The two terms correspond to two time orderings of the two
three-leg vertices, and the ordering is uniquely determined
by the sign of $p_{1'}^+ - p_1^+$.
Using conservation of momentum components $+$ and $\perp$
in the vertices, this factor can be simplified to
\beq
r_{12.1'2'}
\es
  f_{1, t_r} \, f_{2, t_r}
\ ,
\eeq
with
\beq
f_{i, t}
\es
\exp\left[
- t \left( \frac{ m_g^2 - q_i^2 }{ q_i^+ } \right)^2
\right] ,
\label{eq:f_it}
\eeq
where 
\beq
q_1^\mu
\es
p_{1'}^\mu - p_1^\mu
\ ,
\\
q_2^\mu
\es
p_2^\mu - p_{2'}^\mu
\ .
\eeq
Due to momentum conservation, $q_1^+ = q_2^+ = q^+$,
$q_1^\perp = q_2^\perp = q^\perp$.

Now, we write the effective Hamiltonian up to second order in $g_t$,
\beq
H_t
\es
  H_{tf}
+ g_t H_{t\,1}
+ g_t^2 H_{t\,2}
\ .
\label{eq:Ht}
\eeq
The  term $H_{tf}$ is the free Hamiltonian,
\beq
H_{tf}
\es
  \int_1 p_1^- b_{t\,1}^\dag b_{t\,1}
+ \int_2 p_2^- d_{t\,2}^\dag d_{t\,2}
+ \int_3 p_3^- a_{t\,3}^\dag a_{t\,3}
\ . \quad
\eeq
The first-order interaction Hamiltonian is,
\begin{widetext}

\beq
H_{t\,1}
\es
  \int_{131'}
  j_1^\mu
  \ t^3_{11'}
  \ f_{1, t+t_r}
  \ b_{t\,1}^\dag
  \left(
    \tdelta_{13.1'} \, \varepsilon_{3\mu}^* \, a_{t\,3}^\dag
  + \tdelta_{1'3.1} \, \varepsilon_{3\mu} \, a_{t\,3}
  \right) b_{t\,1'}
\nm
  \int_{232'}
  j_2^\mu
  \ t^3_{2'2}
  \ f_{2, t+t_r}
  \ d_{t\,2}^\dag
  \left(
    \tdelta_{23.2'} \, \varepsilon_{3\mu}^* \, a_{t\,3}^\dag
  + \tdelta_{2'3.2} \, \varepsilon_{3\mu} \, a_{t\,3}
  \right) d_{t\,2'}
\ ,
\label{eq:Ht1}
\eeq
\end{widetext}
with $j_1^\mu = \bar u_1 \gamma^\mu u_{1'}$, $j_2^\mu =
\bar v_{2'} \gamma^\mu v_2 = \bar u_2 \gamma^\mu u_{2'}$,
$t^3_{11'} = \chi_{c_1}^\dag T^{c_3} \chi_{c_{1'}}$,
and $t^3_{2'2} = \chi_{c_{2'}}^\dag T^{c_3} \chi_{c_2}$,
where $T^{c_3}$ is half of the Gell-Mann matrix $\lambda^{c_3}$,
$c_3 = 1, 2, \dots, 8$, and $\chi_{c_i} = ( \delta_{1,c_i},
\delta_{2,c_i}, \delta_{3,c_i} )^T$ is the color vector of
quark $i$, $c_i = 1, 2, 3$. This interaction is represented
diagramatically in Fig.~\ref{fig:threeleg}. Note that
$f_{i,t + t_r} = f_{i,t} f_{i,t_r}$, where $f_{i,t_r}$
comes from regularization, while $f_{i,t}$ is a result of
solving Eq.~(\ref{eq:RGPEPeq1}). The fact that they combine
into $f_{i,t + t_r}$ is the motivation behind our choice of
regularization, Eq.~(\ref{eq:regulator}). Whenever
FF energy changes in the interaction by more than $t^{-1/2}$
the form factor $f_{i,t}$ becomes very small, manifesting 
the narrowness of $H_t$.

The second-order interaction Hamiltonian contains the quark-antiquark
interaction term and the quark and antiquark self-interaction terms,
\beq
H_{t \, 2}
\es
H_{Ut}
+ H_{\delta m}
\ .
\label{eq:Ht2}
\eeq
The quark-antiquark interaction term is
\beq
H_{Ut}
\es
- \int_{121'2'} \tdelta_{12.1'2'}
  \, U_{t \, 12.1'2'}
  \, t^a_{11'} t^a_{2'2}
  \ b_{t\,1}^\dag \, d_{t\,2}^\dag \, d_{t\,2'} \, b_{t\,1'}
\ ,
\label{eq:HUt}
\nn
\eeq
where the color superscript $a$ is summed over and the interaction 
kernel $U_{t\,12.1'2'}$ comprises three terms,
\beq
U_{t\,12.1'2'}
\es
  U_C
+ U_H
+ U_X
\ ,
\label{eq:kernelUt}
\eeq
\beq
U_C
\es
  f_{1, t_r} f_{2, t_r} \,
  g_{\mu\nu} j_1^\mu j_2^\nu
  \, f_t
  \, \cF
\ ,
\label{eq:UC}
\\
U_H
\es
- f_{1, t + t_r} f_{2, t + t_r}
  \left(
    \frac{ q_1^2 + q_2^2 }{ 2 (q^+)^2 } \, j_1^+ j_2^+
  + g_{\mu\nu} j_1^\mu j_2^\nu
  \right)
  \cF , \ \ \quad
\\
U_X
\es
   f_t \, f_{1, t_r} f_{2, t_r} \,
  \frac{ j_1^+ j_2^+ }{ (q^+)^2 }
  \left( 1 + \frac{ q_1^2 + q_2^2 }{ 2 } \, \cF \right)
- f_t \, X
\ ,
\label{eq:UX}
\eeq
with
\beq
\cF
\es
\frac{1}{2}
\left(
  \frac{1}{m_g^2 - q_1^2}
+ \frac{1}{m_g^2 - q_2^2}
\right) ,
\eeq
and
\beq
f_t
\es
\exp\left[
- t \left( \frac{\cM_{12}^2 - \cM_{1'2'}^2}{p_1^+ + p_2^+} \right)^2
\right]
\\
\es
\exp\left[
- t \left( \frac{q_2^2 - q_1^2}{p_{3}^+} \right)^2
\right] .
\eeq
The kernel is illustrated in Fig.~\ref{fig:exchange}. Subscripts ``$12.1'2'$,''
which indicate dependence of the kernel on quantum numbers of particles
$1$, $2$, $1'$, and $2'$ are dropped for $U_C$, $U_H$, and $U_X$ to simplify notation.
Details of derivation of $H_{Ut}$ are given in App.~\ref{sec:exchDetail}.
$H_{Ut}$ does not include the terms in which the initial quark-antiquark pair
annihilates into an octet of gluons and then is recreated from the gluons. The octet 
terms are not produced here because they yield zero acting on the color-singlet quark-antiquark states whose dynamics is the main focus of this article. 
The symbol $X$, introduced in Eq.~(\ref{eq:UX}), denotes the contribution of the gluon-exchange counter term, discussed in Sec.~\ref{sec:GExchCT}. The self-interaction term $H_{\delta m}$ in Eq.~(\ref{eq:Ht2}) is discussed in Sec.~\ref{sec:massCT}.

\subsection{Self-interaction counter term\label{sec:massCT}}

Second-order quark self-interaction terms result from successive
action of two 1st-order Hamiltonian interaction terms, in accordance
with Eq.~(\ref{eq:RGPEPsol2nd}) that is illustrated in Fig.~\ref{fig:mass},
\beq
H_{\delta m}
\es
  \int_1 \frac{ \delta m^2_{1t} }{ p_1^+ }
  \, b_{t\,1}^\dag b_{t\,1}
+ \int_2 \frac{ \delta m^2_{2t} }{ p_2^+ }
  \, d_{t\,2}^\dag d_{t\,2}
\ .
\eeq
Details of computing $H_{\delta m}$ are in App.~\ref{sec:massDetail}.
The self-interaction shifts the free quark mass squared, $m_i^2$
in $H_t$, by $g_t^2 \delta m^2_{it}$, where
\beq
\delta m^2_{it}
\es
\delta m_{iX}^2
+ I_i(t + t_r, m_g)
- I_i(t_r, m_g)
\ ,
\label{eq:effectiveMass}
\eeq
with
\beq
I_i(t, m_g)
\es
C_F
\sum_{\sigma_{\tilde i}, \sigma_3}
\int\frac{ d^2 k_{3 \tilde i} d x_{3 \tilde i} }
         { 16\pi^3 x_{3 \tilde i} x_{\tilde i 3} }
\nt 
\frac{ f_{\tilde i 3.i, t}^2 }
       { \cM^2_{\tilde i 3} - m_i^2 }
\ \bar u_i \slashed\varepsilon_3 u_{\tilde i}
  \bar u_{\tilde i} \slashed\varepsilon_3^* u_i
\ ,
\eeq
and $i = 1, 2$, and $\tilde i = \tilde 1, \tilde 2$,
respectively, see Fig.~\ref{fig:mass}. The integrals 
$I_i(t+t_r, m_g)$ are finite for any finite $t > 0$, 
but $I_i(t_r, m_g)$ depends on $t_r$ in a divergent way,

\begin{widetext}
    
\beq
I_i(t_r, m_g)
\es
\frac{C_F}{16\pi^2}
\left\{
  p_i^+
  \sqrt{\frac{\pi}{2 t_r}}
  \left[
    \log\left( \frac{p_i^{+2}}{8 m_g^4 t_r} \right)
  - \frac{7}{2}
  - \gamma
  \right]
+ \frac{1}{2} m_g^2 \log^2\left( \frac{p_i^{+2}}{2 m_i^4 t_r} \right)
\right.
\nn &&
- \left(
    3 m_i^2
  + 4 m_g^2 \log\frac{m_g}{m_i}
  + 3 m_i m_g
  - \frac{3}{2} m_g^2
  + \gamma m_g^2
  \right)
  \log\left( \frac{p_i^{+2}}{2 m_i^4 t_r} \right)
\nn &&
\left.
- 3 m_i^2
+ 3 \gamma m_i^2
\phantom{\sqrt{\frac{}{}}}\hspace{-1em}
\right\}
+ o\left( 1 \right) ,
\label{eq:massCT}
\eeq
\end{widetext}
where $\gamma \approx 0.577$ is the Euler-Mascheroni constant.
The symbol $o\left( 1 \right)$ denotes the terms that tend to 0 when $t_r \to 0$ and
then $m_g \to 0$. One needs the counter term $\delta m_{iX}^2$ 
to remove the divergent part of $I_i(t_r, m_g)$. The finite part
of the counter term is discussed below.

\begin{widetext}

\begin{figure}
 \includegraphics[scale=0.35]{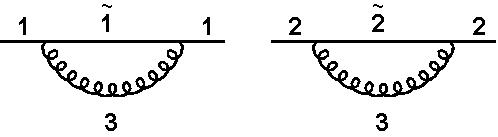}
 \caption{Second-order quark and antiquark self-interaction terms 
 resulting from the product of two first-order interaction terms. \label{fig:mass}}
\end{figure}

\begin{figure}
\includegraphics[scale=0.35]{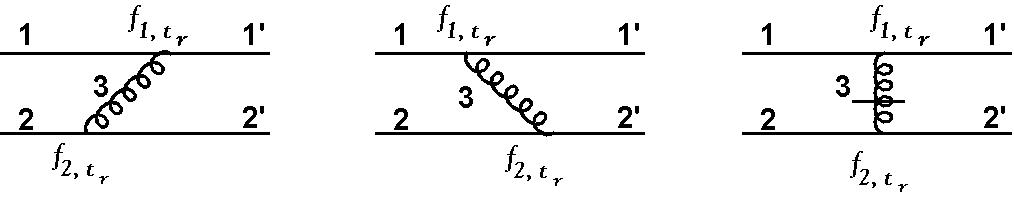}
\caption{Graphs left and middle illustrate the second-order gluon-exchange terms resulting from the product of two first-order interaction terms. 
Right graph illustrates the effective second-order instantaneous 
interaction that results from the unitary rotation of the instantaneous term in the canonical Hamiltonian, see Eq.~(\ref{eq:defcHt}). \label{fig:exchange}}
\end{figure}

\end{widetext}

We define the mass counter term to have the form which removes ``$1$''
in ``$1 - f^2$'' in the Hamiltonian with finite $t$, see Eq.~(\ref{eq:RGPEPsol2ndMass}), by which we mean that the counter term is,
\beq
H_{t_r}^\textrm{mass CT}
\es
  g^2 \int_1 \frac{\delta m_{1X}^2}{p_1^+} b_1^\dag b_1
+ g^2 \int_2 \frac{\delta m_{2X}^2}{p_2^+} d_2^\dag d_2
\ ,
\label{eq:massCTHtr}
\eeq
where
\beq
\delta m_{iX}^2
\es
I_i(t_r, m_g)
\ .
\label{eq:CT}
\eeq
On the one hand, this definition is motivated by the results 
it leads to in our computations of masses of heavy quarkonia~\cite{QbarQ,Serafin_Baryons,Serafin:2019vuk}.
Namely, the singlet quarkonium eigenvalue problem takes a 
simple and phenomenologically reasonable form. At the same 
time the single quark mass eigenvalue tends to infinity 
when $m_g \to 0$, see below. On the other hand, this counter
term removes the ultraviolet divergence from the quark 
self-interaction in the way that is analogous to how the 
electron self-interaction counter term is defined in the
FF Hamiltonian of QED. Our definition of the counter term is 
also compatible with the coupling coherence, which in this 
case implies $\lim_{t \to \infty} \delta m_{it}^2 = 0$, 
see page 66 in Ref.~\cite{Perry:1994mv}. At the current, crude-approximation 
stage of the theory development, the authors find the above 
reasons sufficient for adopting this choice of the quark 
self-interaction counter term including its finite part.

\subsection{Gluon exchange counterterm\label{sec:GExchCT}}

The quark-antiquark interaction term $H_{Ut}$ does not contain 
any loops. However, it leads to the divergent regularization
dependence due to the factor $1/(q^+)^2$ in $U_H$ and $U_X$. As 
long as $t > 0$, $U_H$ is regulated by $f_{1, t + t_r}
f_{2, t + t_r}$. These factors vanish exponentially fast when
$q^+ \to 0$, {\it cf.} Eq.~(\ref{eq:f_it}).
For finite $t > 0$ one can remove the regularization dependence
by setting $t_r = 0$. However, in $U_X$ there is only the 
regulating factor $f_{1, t_r} f_{2, t_r}$, which goes to $1$ 
when $t_r \to 0$. The factor $f_t$ in front of $U_X$ does not 
regulate the singularity when $q^+ \to 0$. Moreover,
\beq
  \frac{ j_1^+ j_2^+ }{ (q^+)^2 }
  \left( 1 + \frac{ q_1^2 + q_2^2 }{ 2 } \, \cF \right)
\es
  \frac{ j_1^+ j_2^+ }{ (q^+)^2 }
  \frac{m_g^2}{m_g^2 - q_1^2}
+ O\left( \frac{1}{q^+} \right) .
\nn
\eeq
Therefore, matrix elements of $H_{Ut}$ diverge for $t_r \to 0$
whenever $q^+ \to 0$.  More precisely, in the vicinity of $q^+ = 0$,
$-q_1^2 \approx \Delta k^2 = (k_{12}^\perp - k_{1'2'}^\perp)^2$, and
the regulator $f_{1, t_r} f_{2, t_r} \approx e^{-2 t_r \frac{(\Delta k^2 + m_g^2)^2}{(q^+)^2}}$.
Integrating $f_{1, t_r} f_{2, t_r}/(q^+)^2$ over $q^+$ gives
$\sqrt{\frac{\pi}{2 t_r \left(\Delta k^2 + m_g^2\right)^2}}$
as the part divergent when $t_r \to 0$. Moreover, $m_g^2/(m_g^2 - q_1^2)
\approx m_g^2/(m_g^2 + \Delta k^2)$. Hence, to counter the
divergence we add the gluon exchange counter term, whose kernel is
\beq
X
\es
\delta(p_{1'}^+ - p_{1}^+)
\, j_{1}^+ j_{2}^+
\, \frac{m_g^2}{(\Delta k^2 + m_g^2)^2}
\, \sqrt{\frac{\pi}{2 t_r}}
\ ,
\label{eq:exchCT}
\eeq
Proper demonstation of cancellation of divergences
is presented in App.~\ref{sec:gluonExchCounterterm}.
Since $m_g^2 (\Delta k^2 + m_g^2)^{-2}$ tends to a
two-dimensional Dirac $\delta$-function of the transverse 
momentum, the counter term is nonzero even in the limit
$m_g \to 0$. It becomes diagonal in momentum and spin. 
We remind the reader the limit $t_r \to 0$ is performed 
before we consider the limit $m_g \to 0$. The latter 
is discussed in Sec.~\ref{sec:quarkonium}.

\subsection{Renormalized Hamiltonian}

With both mass and exchange counter-terms included, one
can remove the ultraviolet regulator, {\it i.e.}, one can take
the limit $t_r \to 0$. It is easily done in $H_{t\,1}$,
$U_C$, and $U_H$. The renormalized mass terms contain 
$m_i^2 + g_t^2 \delta m_{it}^2 = m_i^2 + g_t^2 I_i(t, m_g)$. 
The limit of $U_X$ when $t_r \to 0$ is described in App.~\ref{sec:gluonExchCounterterm}.

\section{Quarkonium in heavy-flavor QCD with gluon mass
ansatz\label{sec:quarkonium}}

In this section, we derive the quarkonium eigenvalue equation using expansion in powers of the coupling constant up to second order. Working at so low order of the expansion, we have to pay a price for 
not knowing what comes out from the non-Abelian interactions of gluons in orders higher than 2nd. In particular, these interactions prevent 
gluons from behaving like photons in QED. Also, the coupling constant 
$g_t$ needs to be extrapolated to values larger than the charge 
$e$ in QED. We assume that the emission and absorption of effective 
gluons by effective quarks and antiquarks is blocked for sizable 
$t$. We model the non-Abelian blocking by introducing the gluon-mass 
$m_G$ for the effective gluons. The resulting eigenvalue equations 
for quarkonia and single quarks are then obtained keeping $m_g > 0$. 
Subsequently, we discuss these equations in the limit $m_g \to 0$.

\subsection{Effective Hamiltonian}

Our description of heavy quarkonium closely follows Ref.~\cite{QbarQ}. Here we focus on the main steps, {\it cf.}~\cite{Glazek:2021vnw}. We consider the quarkonium eigenvalue problem assuming that a single 
quark-antiquark pair gives the dominant contribution. Other Fock sectors are included using expansion in powers of $g_t$. Up to the second order, we need two Fock sectors: the leading quark-antiquark sector $Q \bar Q$ and the quark-antiquark-gluon sector $Q \bar Q G$. The large letter $G$ signifies the effective gluons whose mass is assigned a hypothetical value $m_G$. Namely, we  modify the Hamiltonian 
limited to  $Q \bar Q$ and $Q \bar Q G$ by adding to it a nonperturbative gluon mass term,
\beq
\hat m_G^2
\es
\int_{123}
\, \frac{m_G^2}{p_3^+}
\ b_{t\,1}^\dag d_{t\,2}^\dag a_{t\,3}^\dag
\ket{0}\bra{0}
a_{t\,3} d_{t\,2} b_{t\,1}
\ .
\eeq
This operator acts only in the $Q \bar Q G$ sector. In principle, $m_G$ could be a multiple of $\Lambda_{\rm QCD}$ in the RGPEP scheme and hence not expandable in powers of $g_t$. More about our gluon mass ansatz can be found in Refs.~\cite{QbarQ,QQ1}. 

Perturbative computation of the effective Hamiltonian~\cite{Wilson1970} in the $Q \bar Q$ sector yields $H_\mathrm{eff}$ whose matrix elements are
\begin{widetext}
    
\beq
\bra{L} H_\mathrm{eff} \ket{R}
\es
\bra{L} \left[
  H_{11} 
+ \frac{1}{2} H_{12}
  \left(
    \frac{1}{E_L - H_{22} - \hat m_G^2}
  + \frac{1}{E_R - H_{22} - \hat m_G^2}
  \right) H_{21}
\right] \ket{R}
\ ,
\label{eq:HeffDEFQQbarSinglet}
\eeq

\end{widetext}
where $H_{ij} = P_i H_t P_j$, with $P_1$ and $P_2$ the projection
operators onto the $Q \bar Q$ and $Q \bar Q G$ sectors,
respectively. We keep only the free part of $H_{22}$ in the denominators, because other terms in $H_{22}$ are of order 
$g^2$ and contribute terms in $H_\mathrm{eff}$ of at least 4th order. The states
\beq
\ket{L}
\es
\int_{12}
P_L^+ \tdelta_{12.P_L}
\frac{\delta_{c_1,c_2}}{\sqrt{N_c}}
\, \psi_L(1,2)
\ b_{t\,1}^\dag d_{t\,2}^\dag \ket{0}
\ ,
\label{eq:ketL}
\\
\ket{R}
\es
\int_{1'2'}
P_R^+ \tdelta_{1'2'.P_R}
\frac{\delta_{c_{1'},c_{2'}}}{\sqrt{N_c}}
\, \psi_R(1',2')
\ b_{t\,1'}^\dag d_{t\,2'}^\dag \ket{0}
\ ,
\label{eq:ketR}
\eeq
are eigenstates of the free part of $H_{11}$ with the eigenvalues
$E_L = p_1^- + p_2^- = [\cM_{12}^2 + (P_L^\perp)^2]/P_L^+$
and $E_R = p_{1'}^- + p_{2'}^- = [\cM_{1'2'}^2 + (P_R^\perp)^2]/P_R^+$,
respectively. Due to momentum conservation only matrix elements
with $P_L = P_R \equiv P$ are nonzero.  The evaluation of the matrix
elements gives
\beq
\bra{L} H_\mathrm{eff} \ket{R}
\es
P^+ \tdelta_{P_L.P_R}
\sum_{\sigma_1, \sigma_2}
\int[12]
P_L^+ \tdelta_{12.P_L}
\, \psi_L^*(1,2)
\ \left(\cE\psi_R\right)(1,2)
\ ,
\eeq
where
\beq
\left(\cE\psi_R\right)(1,2)
\es
\left(
  \frac{m_1^2 + \mathscr{M}_1^2 + p_1^{\perp 2}}{p_1^+}
+ \frac{m_2^2 + \mathscr{M}_2^2 + p_2^{\perp 2}}{p_2^+}
\right)
\psi_R(1,2)
\nm
  C_F g_t^2
  \sum_{\sigma_{1'}, \sigma_{2'}}
  \int[1'2']
  \, \tilde\delta_{12.1'2'}
  \, \tilde U_{t\,12.1'2'}
  \, \psi_R(1',2') \ .
\label{eq:Ecolorsinglet}
\eeq
The momentum integration measure for two particles,
$1$ and $2$, is,
\beq
[12]
\es
\frac{dp_1^+ d^2 p_1^\perp}{16\pi^3 p_1^+}
\frac{dp_2^+ d^2 p_2^\perp}{16\pi^3 p_2^+}
\ ,
\eeq
and analogously $[1'2']$ for particles $1'$ and $2'$.
The self-interaction terms are
\beq
\mathscr{M}_i^2
\es
  C_F g_t^2
  \sum_{\sigma_{\tilde i}, \sigma_3}
  \int\frac{ d^2 k_{3 \tilde i} dx_{3 \tilde i} }
           { 16 \pi^3 x_{3 \tilde i} x_{\tilde i 3} }
  \,
  \frac{m_G^2}{x_{3 \tilde i}}
  \frac{ f_{\tilde i3.i, \, t}^2 }
       { \left( \cM_{\tilde i 3}^2 + \frac{m_G^2}{x_{3 \tilde i}} - m_i^2 \right)
         \left( \cM_{\tilde i 3}^2 - m_i^2 \right) }
  \,
  \bar u_i \slashed\varepsilon_3 u_{\tilde i}
  \bar u_{\tilde i} \slashed\varepsilon_3^* u_i
\ ,
\label{eq:msMi2}
\eeq
and the effective $Q \bar Q$ interaction kernel, which one can call the $Q \bar Q$ potential, is
\beq
\tilde U_{t\,12.1'2'}
\es
  U_C
+ \tilde U_H
+ U_X
\ ,
\eeq
where $U_C$ and $U_X$ are given in Eqs.~(\ref{eq:UC}) and
(\ref{eq:UX}), respectively, and
\beq
\tilde U_H
\es
  f_{1, t} f_{2, t} \,
  \left(
    \frac{ q_1^2 + q_2^2 }{ 2 (q^+)^2 } \, j_1^+ j_2^+
  + g_{\mu\nu} j_1^\mu j_2^\nu
  \right)
  \left[
    \frac{1}{2}
    \left(
      \frac{1}{ m_G^2 + m_g^2 - q_1^2 }
    + \frac{1}{ m_G^2 + m_g^2 - q_2^2 }
    \right)
  - \cF
  \right] .
\nn
\label{eq:tildeV_H}
\eeq

\subsection{Color-singlet quarkonium}

Consider the eigenvalue problem,
\beq
\left(\cE\psi\right)(1,2)
\es
\frac{M^2 + P^{\perp 2}}{P^+} \, \psi(1,2)
\ ,
\label{eq:eve}
\eeq
where $M^2$ is the mass of the bound state. There are two types of terms
in $\cE$, self-interactions and potential terms. Both diverge like
$|\log m_g|$ when $m_g \to 0$. However, the self-interactions diverge to the positive infinity, while the potential diverges to the negative infinity.
In order to isolate the divergence of the potential terms we
rewrite 
\beq
\left(\cE\psi\right)(1,2)
\es
\left(
  \frac{m_1^2 + \mathscr{M}_1^2 + p_1^{\perp 2}}{p_1^+}
+ \frac{m_2^2 + \mathscr{M}_2^2 + p_2^{\perp 2}}{p_2^+}
- \frac{\Delta}{p_1^+ + p_2^+}
\right)
\psi(1,2)
\nm
  C_F g_t^2
  \sum_{\sigma_{1'}, \sigma_{2'}}
  \int[1'2']
  \, \tilde\delta_{12.1'2'}
  \, \tilde U_H
  \left[ \psi(1',2') - \delta_{\sigma_1, \sigma_{1'}} \delta_{\sigma_2, \sigma_{2'}} \psi(1,2) \right]
\nm
  C_F g_t^2
  \sum_{\sigma_{1'}, \sigma_{2'}}
  \int[1'2']
  \, \tilde\delta_{12.1'2'}
  \left( U_C + U_X \right)
  \, \psi(1',2')
\ ,
\label{eq:cErewrite}
\eeq
where
\beq
\Delta
\es
  C_F g_t^2
  \sum_{\sigma_{1'}, \sigma_{2'}}
  \int[1'2']
  \, (p_1^+ + p_2^+) \, \tilde\delta_{12.1'2'}
  \, \tilde U_H
  \, \delta_{\sigma_1, \sigma_{1'}} \delta_{\sigma_2, \sigma_{2'}}
\label{eq:defDelta} \ ,
\eeq
results from subtracting and adding $\delta_{\sigma_1, \sigma_{1'}} \delta_{\sigma_2, \sigma_{2'}}\psi(1,2)$ to $\psi(1',2')$
under the integral $\int[1'2']$. The most singular part of the integrand appears in the vicinity of $q^+ = q_1^+ = q_2^+ = 0$ 
while $q^\perp = q_1^\perp = q_2^\perp = 0$. The singularity is regulated by $m_g$. Near that singular region the integrand is proportional to $(p_1^+ + p_2^+) f_{1,t} f_{2,t}/(q^+)^2$, where
$f_{1,t} \approx f_{2,t} \approx e^{-t (p_3^-)^2}
= \exp\left\{- t [m_g^2 + q^{\perp 2}]^2/(q^+)^2\right\}$.
Integration over $q^+$ from $-p_1^+$ to $p_2^+$ and over
$q^\perp$ over the whole two-dimensional transverse plane 
gives,
\beq
\Delta
\es
  \frac{C_F g_t^2}{8 \pi^2}
  \log\left( \frac{p_1^+ p_2^+}{8 m_g^4 t} \right)
  \sqrt{\frac{\pi}{2 t}}
  \, (p_1^+ + p_2^+)
+ O(m_g^0) \ ,
\eeq
where $O(m_g^0)$ denotes the terms that are finite in the 
limit $m_g \to 0$. Similarly, the integrand in the self-interaction
of Eq.~(\ref{eq:msMi2}) near $p_3^+ = 0$ is approximately
$p_i^+ f_{i,t}^2/(p_3^+)^2$, where $f_{i,t} \approx
\exp\left\{- t [m_g^2 + (p_3^\perp)^2]^2/(p_3^+)^2\right\}$
and $k_{3\tilde i}^\perp \approx p_3^\perp$.
Integration over $p_3^+$ from 0 to $p_i^+$ and over
$p_3^\perp$ over the whole two-dimensional plane gives,
\beq
\mathscr{M}_i^2
\es
  \frac{C_F g_t^2}{16 \pi^2}
  \log\left( \frac{p_i^{+2}}{8 m_g^4 t} \right)
  \sqrt{\frac{\pi}{2 t}}
  \, p_i^+
+ O(m_g^0)
\ .
\eeq
The logarithms of the gluon mass $m_g$ obtained above 
cancel out in
\beq
  \frac{\mathscr{M}_1^2}{p_1^+}
+ \frac{\mathscr{M}_2^2}{p_2^+}
- \frac{\Delta}{p_1^+ + p_2^+}
\es
  O(m_g^0)
\ .
\label{eq:cancellation}
\eeq
The eigenvalue problem has a finite limit for $m_g \to 0$. 
However, the quarkonium mass eigenvalue $M^2$ depends on $P^+ =
p_1^+ + p_2^+$ of the state. The dependence comes mainly from 
the potential produced by the second term on the right-hand 
side of Eq.~(\ref{eq:cErewrite}). That potential is confining, 
{\it cf.} Eq.~(\ref{eq:Vconf}) below. For small distances $r$ 
between $Q$ and $\bar Q$ it behaves as $r^2$. The associated 
oscillator frequency $\omega$ is proportional to 
$(t/P^{+2})^{-3/4}$. Therefore, it is natural to set the longitudinal momentum scale $\cP$ of Eq.~(\ref{eq:tlambda}) to the quarkonium momentum, $\cP = P^+$, which implies $\omega \sim \lambda^3$,
see Sec.~\ref{sec:nonrel}.

\subsection{Color-octet quarkonium and eigenquarks}

For color-octet quark-antiquark states one can proceed by
the same steps as for the color-singlet case. The quark-antiquark
annihilation interaction needs to be included. Its contribution 
is finite in the limit $m_g \to 0$. Now, the color-octet wave 
functions of the states $\ket{L}$ and $\ket{R}$ lead to the 
different color factors in the potential term of the octet 
eigenvalue equation, $(2 N_c)^{-1} = 1/6$ instead of $-C_F = -4/3$.
The self-interactions $\mathscr{M}_i^2$ do not depend on the color 
wave function of quarkonium. Accordingly, the term $\Delta$ 
includes the factor $-(2 N_c)^{-1}$. Therefore, instead of 
Eq.~(\ref{eq:cancellation}), we obtain for the octet states
\beq
  \frac{\mathscr{M}_1^2}{p_1^+}
+ \frac{\mathscr{M}_2^2}{p_2^+}
- \frac{\Delta}{p_1^+ + p_2^+}
\es
  N_c \, \frac{g_t^2}{16 \pi^2}
  \log\left(\frac{p_1^+ p_2^+}{2 m_g^4 t}\right)
  \sqrt{\frac{\pi}{2 t}}
+ O(m_g^0)
\ .
\eeq 
The cancellation of $|\log m_g|$ is absent. The expectation 
value of $H_\mathrm{eff}$ in the color-octet states diverges 
to plus infinity in the limit $m_g \to 0$.

Similar non-cancellation of $|\log m_g|$ appears in the eigenvalue equations for states with quantum numbers of a single quark, which 
we for brevity call eigenquarks. The eigenquark mass eigenvalue  
$M_Q$ diverges to plus infinity when $m_g \to 0$. Assuming the same $m_G$ as in the quarkonium case, we obtain
\beq
M_Q^2
\es
  m_1^2
+ C_F g_t^2
  \sum_{\sigma_{\tilde 1}, \sigma_3}
  \int\frac{d^2k_{3 \tilde 1} dx_{3 \tilde 1}}{16\pi^3 x_{3 \tilde 1} x_{\tilde 13}}
  \, \frac{m_G^2}{x_{3 \tilde 1}}
  \frac{ f_{\tilde 13.1, t}^2 }
       { \left( \cM_{\tilde 13}^2 + \frac{m_G^2}{x_{3 \tilde 1}} - m_1^2 \right)
         \left( \cM_{\tilde 13}^2 - m_1^2 \right) }
  \,
  \bar u_1 \slashed\varepsilon_3 u_{\tilde 1}
  \bar u_{\tilde 1} \slashed\varepsilon_3^* u_1
\\
\es
  m_1^2
+ P^+
  \frac{C_F g_t^2}{16 \pi^2}
  \sqrt{\frac{\pi}{2 t}}
  \left[
    \log\left( \frac{P^{+2}}{2 m_g^4 t} \right)
  - 2
  + \frac{1}{\sqrt{\pi}}
    \int_0^\infty ds' \,
    e^{-s'^2}
    f\left( s', \frac{m'^2}{P^+},
    \frac{m_G'^2}{P^+} \right)
  \right]
+ o(m_g^0)
\ , \nn 
\eeq
where $P^+$ is the longitudinal momentum of the eigenquark state,
$m'^2 = \sqrt{2 t} m_1^2$, and $m_G'^2 = \sqrt{2 t} m_G^2$, and
\beq
f\left( a, b, c \right)
\es
  4 \log\left(\frac{a^2}{a + b}\right)
+ \frac{2 c}{a + b}
\nm
  4 c
  \left(
    \frac{1}{c}
  + \frac{1}{a}
  + \frac{b + c/2}{a^2}
  \right)
  \log\left(
    1
  + \frac{a^2}{c \left(a + b\right)}
  \right) .
\eeq
Function $f(a,b,c)$ should not be confused with any
of the form factors. Terms $o(m_g^0)$ vanish when $m_g \to 0$.

\section{Nonrelativistic approximation of the effective potential}
\label{sec:nonrel}

The eigenvalue equation~(\ref{eq:eve}) has interesting properties.
We exhibit them using the nonrelativistic (NR) limit in which the quark masses are considered very large. In that limit quarks have typical longitudinal momentum fractions $x_1 \approx x_{10}= m_1/(m_1 + m_2)$ and $x_2 \approx x_{20} = m_2/(m_1 + m_2)$. Information about the state resides in the wave function dependence on the deviation of $x_1$ from $x_{10}$ and the quarks relative motion in the transverse directions. We introduce the quark three-dimensional relative momentum $\vec k = (k^x, k^y, k^z)$, 
\beq
\vec k
\es
\sqrt{\frac{m_1 m_2}{x_1 x_2}}
\left(
\frac{k_{12}^x}{m_1 + m_2},\ 
\frac{k_{12}^y}{m_1 + m_2},\ 
x_1 - x_{10}
\right) ,
\eeq
which generalizes the definition of $\vec k$ in Ref.~\cite{QbarQ} to the case of $m_1 \neq m_2$, both masses being large. The potential is a function of $\vec k$ and $\vec k'$. The NR limit is obtained assuming that $|\vec k|, |\vec k'| \ll m_1 + m_2$ and keeping only the leading terms. After multiplying Eq.~(\ref{eq:eve}) by $P^+/(2 m_1 + 2 m_2)$, the NR limit of the eigenvalue equation reads
\beq
\left(
  \frac{\vec k^2}{2 \mu}
+ \frac{\mathscr{M}_1^2}{2 m_1}
+ \frac{\mathscr{M}_2^2}{2 m_2}
\right)
\psi_{\sigma_1, \sigma_2}( \vec k )
- C_F g_t^2
  \int\frac{d^3 k'}{(2\pi)^3}
  \left(
    V_C
  + V_H
  + V_X
  \right)
  \, \psi_{\sigma_1, \sigma_2}( \vec k' )
\es
E \, \psi_{\sigma_1, \sigma_2}( \vec k ) , 
\nn
\label{eq:QQbarEVENR1}
\eeq
where the reduced mass $\mu = m_1 m_2 / (m_1 + m_2)$. The potentials
$V_C$, $V_H$, and $V_X$ are the nonrelativistic approximations
of $U_C/(4 m_1 m_2)$, $\tilde U_H/(4 m_1 m_2)$,
and $U_X/(4 m_1 m_2)$, respectively. We obtain
\beq
V_C
\es
  \frac{ f_t }{ \vec q\,^2 + m_g^2 }
\ ,
\\
V_H
\es
  f_{1, t} f_{2, t}
  \left(
    \frac{ 1 }{ (q^z)^2 }
  - \frac{ 1 }{ \vec q\,^2 }
  \right)
  \frac{ \vec q\,^2 }{ \vec q\,^2 + m_g^2 } \,
  \frac{ m_G^2 }{ m_G^2 + m_g^2 + \vec q\,^2 }
\ ,
\\
V_X
\es
0
\ ,
\eeq
where $\vec q = \vec k' - \vec k$. $V_X$ is set to 0 because 
it is suppressed by the inverse of $\mu^2$ in comparison to 
$V_C$ and $V_H$, see Eq.~(\ref{eq:VXNR}). The eigenvalue $E =
[M^2 - (m_1 + m_2)^2]/[2(m_1 + m_2)] \approx M - (m_1 + m_2)$,
since $M \approx m_1 + m_2$. The RGPEP form factors are
\beq
f_{i, t}
\es
\exp\left(
- \frac{t (m_1 + m_2)^2}{(p_1^+ + p_2^+)^2}
  \frac{ \left( \vec q\,^2 + m_g^2 \right)^2 }{ (q^z)^2 }
\right) ,
\label{eq:NRfi}
\\
f_t
\es
\exp\left(
- \frac{t (m_1 + m_2)^4
  \left( \vec k^2 - \vec k'^2 \right)^2}{
  (p_1^+ + p_2^+)^2 m_1^2 m_2^2}
\right) .
\eeq
The NR approximation for the self-interaction terms $\mathscr{M}_i^2$
is obtained using variables $\tilde q^{x,y} = k_{3 \tilde i}^{x,y}$, $\tilde q^z = x_{3 \tilde i} m_i$ that appear in the form factors. 
We find
\beq
\mathscr{M}_i^2
\es
  m_i
  C_F g_t^2
  \int\frac{ d^3 \tilde q }
           { (2 \pi)^3 }
  \, f_{i, t}^2
  \left( \frac{1}{(\tilde q^z)^2} - \frac{1}{\tilde q^2} \right)
  \frac{ \tilde q^2 }{ \tilde q^2 + m_g^2 }
  \frac{ m_G^2 }{ m_G^2 + m_g^2 + \tilde q^2 }
\\
\es
  m_i
  C_F g_t^2
  \int\frac{ d^3 q }
           { (2 \pi)^3 }
  \, V_H
\ ,
\eeq
where $f_{i, t}$ is the same as in Eq.~(\ref{eq:NRfi}) except
that $\vec q$ is replaced by $\vec{\tilde q}$. The second
equality holds because in the NR approximation $f_{1, t} = 
f_{2, t}$. Therefore, the self-energy terms $\mathscr{M}_1^2$ and $\mathscr{M}_2^2$ can be combined with the potential term $V_H$. 
The quarkonium eigenvalue equation in the limit $m_g \to 0$ becomes
\beq
  \frac{\vec k^2}{2 \mu} \, \psi( \vec k )
+ \left( V_\mathrm{conf} \, \psi \right)(\vec k)
- \int\frac{d^3 k'}{(2\pi)^3}
  \, \frac{C_F g_t^2 f_t}{\vec q\,^2}
  \, \psi( \vec k' )
\es
E \, \psi( \vec k )
\ ,
\label{eq:QQbarEVENR2}
\eeq
where the spin indices are omitted and, by definition,
\beq
\left( V_\mathrm{conf} \, \psi \right)(\vec k)
\es
\lim_{m_g \to 0^+}
  (- C_F g_t^2)
  \int\frac{d^3 q}{(2\pi)^3}
  \, V_H
  \left[
    \psi( \vec k + \vec q\, )
  - \psi( \vec k )
  \right] .
\eeq
The limit $m_g \to 0$ is well-defined, because the difference
of wave functions regulates $1/(q^z)^2$ in $V_H$. We expand 
$\psi(\vec k + \vec q\,)$ in a Taylor series in $\vec q$. 
Only terms with even powers of $q^x$, $q^y$, and $q^z$ 
contribute to the result of integration over $\vec q$ 
because $V_H$ is even in $\vec q$. Therefore,
\beq
\left( V_\mathrm{conf} \, \psi \right)(\vec k)
\es
{\sum_{n,k}}'
C_{n,k}
\left( \frac{\pd^2}{(\pd k^x)^2}
     + \frac{\pd^2}{(\pd k^y)^2} \right)^n
\left( \frac{\pd}{\pd k^z} \right)^{2k}
\psi(\vec k)
\ ,
\eeq
where ${\sum_{n,k}}'$ is the sum over all nonnegative
$n$ and $k$ with the exception of the $n = k = 0$ term,
and
\beq
C_{n,k}
\es
- \frac{C_F g_t^2}{(2\pi)^2}
  \frac{ 2 }{ (2^n n!)^2 (2k)! }
  \int_0^1 dw
  (1 - w^2)^{n+1}
  w^{2k-2}
  \int_0^\infty d q \,
  \frac{ m_G^2 }{ m_G^2 + q^2 }
  \, q^{2n + 2k}
  \, e^{-\frac{ t' q^2 }{ w^2 }}
\ ,
\eeq
where
\beq
t'
\es
\frac{2 t \left( m_1 + m_2 \right)^2}
{\left( p_1^+ + p_2^+ \right)^2}
\rs
\frac{2 \cP^2 \left( m_1 + m_2 \right)^2}
{\lambda^4 \left( p_1^+ + p_2^+ \right)^2}
\ .
\label{eq:deftp}
\eeq
Coefficients $C_{n,k}$ can be evaluated explicitly
for $m_G^2 \to \infty$,
\beq
\lim_{m_G^2 \to \infty}
C_{n,k}
\es
- \frac{C_F g_t^2}{(2\pi)^2}
  \frac{ \sqrt{\pi} }{ 2^{3n+k+1} }
  \frac{(n+1) t'^{-k-n-\frac{1}{2}}}{(2k+2n)!!(2k+2n+1)(2 k+n)}
  \binom{n + 2k}{n}
\, .
\eeq
The potential in position space is obtained by substituting,
$\frac{\pd^2}{(\pd k^x)^2} + \frac{\pd^2}{(\pd k^y)^2}
\to -\rho^2 = - x^2 - y^2$ and
$\left( \frac{\pd}{\pd k^z} \right)^2 \to -z^2$. Consequently,
\beq
V_\mathrm{conf}(\vec r\,)
\rs
V_\mathrm{conf}(\rho, z)
\es
{\sum_{n,k}}' C_{n,k}
\left( -\rho^2 \right)^n
\left( -z^2 \right)^k .
\label{eq:Vconf}
\eeq
The double summation makes it difficult to obtain the potential
as a function of $\vec r$. However, for small separations between 
the quark and antiquark only the quadratic terms in $\vec r$ count.
Moreover, in the limit $m_G \to \infty$ one obtains $C_{1,0} = 
C_{0,1}$ and the potential is spherically symmetric as a function 
of $\vec r$,
\beq
V_\mathrm{conf}\left( \vec r\, \right)
\es
  \frac{C_F g_t^2}{192 \pi^{3/2} t'^{3/2}} \, \vec r\,^2
+ O\left( \frac{r^4}{t'^{5/2}} \right) .
\eeq
For arbitrary separations we found simple expressions
along some special directions of $\vec r$ in the limt 
$m_G^2 \to \infty$.
\beq
V_\mathrm{conf}(\rho, 0)
\es
\frac{C_F g_t^2}{8 \pi^{3/2}}
\left[
  \frac{2 \sqrt{\pi}}{\rho} \,
  \mathrm{erf}\left(\frac{\rho}{4 \sqrt{t'}}\right)
+ \frac{
    \log\left(\frac{\rho^2}{16 t'}\right)
  + E_1\left(\frac{\rho^2}{16 t'}\right)
  + \gamma
  - 1
  }{\sqrt{t'}}
\right] ,
\\
V_\mathrm{conf}(0, z)
\es
\frac{C_F g_t^2}{16 \pi^{3/2}}
\left[
  \frac{2 \sqrt{\pi}}{z} \,
  \mathrm{erf}\left(\frac{z}{2 \sqrt{t'}}\right)
+ \frac{
    \log\left(\frac{z^2}{4 t'}\right)
  + E_1\left(\frac{z^2}{4 t'}\right)
  + \gamma
  - 2
  }{\sqrt{t'}}
\right] ,
\eeq
where $\mathrm{erf}$ is the error function and $E_1$
is the exponential integral function. For large $\rho$
or $z$ (or small $t'$) the effective potential becomes
logarithmic,
\beq
\label{V1}
V_\mathrm{conf}\left( \rho, 0 \right)
\es
  \frac{C_F g_t^2}{8 \pi^{3/2}}
  \frac{
    \log\left( \frac{\rho^2}{16 t'} \right)
  + \gamma
  - 1
  }{ \sqrt{t'} }
+ \frac{C_F g_t^2}{4 \pi \rho}
+ O\left( \frac{ \sqrt{t'} e^{-\frac{\rho^2}{16 t'}} }{\rho^2} \right)
 ,
\\
\label{V2}
V_\mathrm{conf}( 0, z )
\es
  \frac{C_F g_t^2}{16 \pi^{3/2}}
  \frac{
    \log\left( \frac{z^2}{4 t'} \right)
  + \gamma
  - 2
  }{\sqrt{t'}}
+ \frac{C_F g_t^2}{8 \pi z}
+ O\left( \frac{ t'^{3/2} e^{-\frac{z^2}{4 t'}} }{z^4} \right)
 .
\eeq
Our result agrees at large distances with the logarithmic 
potentials found in Ref.~\cite{Perry:1994mv}, if one sets $\frac{\Lambda^2}{\cP^+} =
\frac{1}{2} \sqrt{\frac{\pi}{2}} \frac{\lambda^2}{m_1 + m_2}
\frac{p_1^+ + p_2^+}{\cP}$, where $\Lambda^2/\cP^+$ is
the $p^-$ cutoff of Ref.~\cite{Perry:1994mv}.

\begin{figure}
   \centering
    \includegraphics{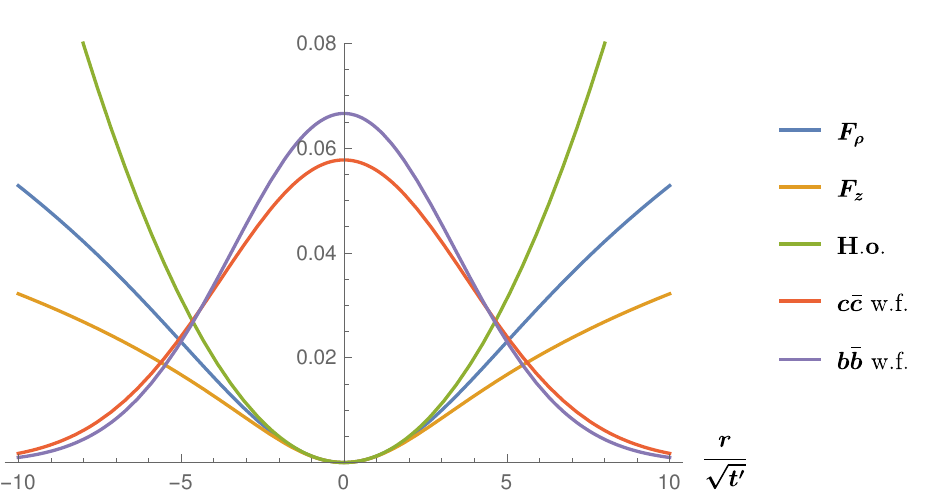}
    \caption{Harmonic oscillator approximation compared to the full potential, as functions of the distance $r$ between quark and antiquark in units of $\sqrt{t'}$, see Eq.~(\ref{eq:deftp}). The ground state wave functions, $c \bar c$ w.f. and $b \bar b$ w.f., are obtained fitting the corresponding spectra. They are plotted for comparison with the potential. The harmonic oscillator H.o. is an extrapolation of the quadratic behavior of the potential near $r \sim 0$. Functions $F_\rho$ and $F_z$ are explained in the text below Eqs.~(\ref{V1}) and (\ref{V2}).}
    \label{fig:HOvsFull}
\end{figure}
Figure~\ref{fig:HOvsFull} shows the accuracy of the harmonic 
oscillator approximation for the potential. We first notice that 
$\frac{\sqrt{t'}}{g_t^2} V_ \mathrm{conf}(\rho, 0) = F_\rho \left( \frac{\rho}{\sqrt{t'}} \right)$ and
$\frac{\sqrt{t'}}{g_t^2} V_\mathrm{conf}(0, z) = F_z \left( \frac{z}{\sqrt{t'}} \right)$,
where the  functions $F_\rho$ and $F_z$ do not depend on the
parameters of the theory. In the harmonic oscillator
approximation, $F_\rho(r/\sqrt{t'}) = F_z(r/\sqrt{t'})
= r^2/144 t' \pi^{3/2}$. Typical separations between quarks
in ground-states of heavy quarkonia can be extracted from
their electromagnetic form factors~\cite{Serafin:2019vuk}. Using
the values for $\lambda$, $g_t$, and the quark masses fitted
in Ref.~\cite{Serafin_Baryons} for charmonium and bottomonium,
we get\footnote{Since Ref.~\cite{Serafin_Baryons} uses a different RGPEP generator, we have to rescale $\lambda$ in order to obtain the same spectroscopy. If we define $\lambda_\mathrm{old}$ to be the value of $\lambda$ used in Ref.~\cite{Serafin_Baryons} and $\lambda_\mathrm{new}$ to be the value of $\lambda$ we are using here, then, in Ref.~\cite{Serafin_Baryons} $t' = (m_1^2 + m_2^2)/\lambda_\mathrm{old}^4$, while from Eq.~(\ref{eq:deftp}) we have here $t' = 2(m_1 + m_2)^2/\lambda_\mathrm{new}^4$ (assuming $\cP = p_1^+ + p_2^+$). Therefore, for equal quark masses, $m_1 = m_2$, we have
$\lambda_\mathrm{new} = \sqrt{2} \lambda_\mathrm{old}$.}:
For bottomonium $\sqrt{t'} \approx 0.37~\text{GeV}^{-1} \approx 0.073$~fm,
while $r_\mathrm{EM} \sim 0.15$~fm, where $r_\mathrm{EM}$ is the radius
extracted from the electromagnetic form factors~\cite{Serafin:2019vuk}.
The relative separation between quark and antiquark is $r \sim 2 r_\mathrm{EM}$.
Therefore, typical $r/\sqrt{t'}$ for bottomonium is 4.1. For charmonium
$\sqrt{t'} \approx 0.55~\text{GeV}^{-1} \approx 0.109$~fm, while
$r_\mathrm{EM} \sim 0.25$~fm. Therefore, typical $r/\sqrt{t'} \sim 4.6$.
Harmonic oscillator potential becomes twice too strong in the z direction
for $r/\sqrt{t'} \approx 5.4$ at which point it is approximately 1.4
times too strong in the transverse direction.

In the excited states, an important weakening of the potential
occurs due to the change of a quadratic behavior over to a 
logarithmic one. In addition, the excited states are likely
to be sensitive to the details of gluon components. Such details
are not accounted for  in any way by the gluon mass ansatz and
the resulting oscillator. To derive these details one needs to
solve the RGPEP equation to higher order than second and include
effective components with more gluons than one, shifting the mass
ansatz to sectors with the maximal number of gluons one includes
in numerical computations.

\section{Conclusion}

Once canonical FF Hamiltonian for heavy-quark QCD in gauge $A^+=0$
is supplied with a small gluon mass $m_g$ and subjected to the RGPEP scale-evolution of second order in a weak-coupling expansion, a simple dynamical picture is obtained in terms of the resulting eigenvalue equations for quarkonia at the scale of quark masses, provided that the emission and absorption of the effective gluons by quarks is blocked by assigning to them a hypothetically large effective gluon mass $m_G$. Quark self-interactions diverge in the limit $m_g \to 0$ but the divergence is canceled by the effective  
quark-antiquark interaction in color-singlet states. In color-octet
states the cancellation does not occur. As a consequence, they cannot have finite masses in that limit.  Single quark mass also cannot be finite. The finite color-singlet quarkonium eigenvalue problem can be further analyzed using the non-relativistic approximation. The effective quark-antiquark potential at small distances $r$ between the quarks includes a Coulomb term and a spherically symmetric oscillator term. The latter turns at large distances into a logarithmic dependence on $r$ with different strengths in the transverse and longitudinal directions, matching the confining potential obtained by Perry using coupling coherence. Previous calculations of white quarkonia masses with such potentials indicate that the effective dynamics is likely to explain the ground and low excited states when the effective quark and gluon dynamics is computed using the RGPEP for heavy quarks more accurately. However, inclusion of light quarks in the dynamics would initially require guessing effective masses for them in a similar way to how it is done 
here for gluons.

Systematic increase in accuracy of quarkonium dynamics may be sought using the RGPEP by computing the running of $H_t$ for heavy quarks in orders higher than 2nd. The actual running of the coupling constant $g_t$ shows up in the quarkonium dynamics first in 4th order. Increasing the order implies inclusion of Fock components with more than one effective gluon in the hadronic states that are described numerically by solving the nonperturbative eigenvalue problem of $H_t$. The ansatz $m_G$ must be shifted to the sector with the highest number of gluons. One may hope that the ansatz is eventually eliminated when the gluon effects are saturated by increasing their number.

Blocking effective gluons from significant involvement in the dynamics of lowest-mass quarkonia using mass $m_G$ may reasonably reflect the effective gluons behavior because the results summarized above do not depend qualitatively on the value of $m_G$ when it exceeds the scale of relative momentum of effective quarks, $p \sim \sqrt{\mu \omega} = \sqrt{m \omega/2}$. Our mass spectrum fit for quarkonia yields for $c \bar c$ that $m_c \sim$ 1.46~GeV, $\omega \sim$ 322~MeV and  $p_{c \bar c} \sim$ 0.5~GeV. For $b \bar b$, we get $m_b \sim$ 4.7~GeV, $\omega_{b \bar b} \sim$ 269~MeV and $p_{b \bar b} \sim$ 0.8~GeV. More accurate computations than ours can provide more precise estimates of $m_G$, if $m_G$ turns out to properly grasp the mechanism of blocking gluons. It is not known yet if a universal $m_G$ may result from the RGPEP evolution that includes some arbitrarily small $m_g$ added to the canonical QCD Hamiltonian at $t=0$.

\section*{Acknowledgement}

K. S. acknowledges support by the Senior Scientist Program
funded by Gansu Province, Grant No. 22JR10KA001, by Gansu
International Collaboration and Talents Recruitment Base
of Particle Physics (2023-2027), by the Chinese Academy
of Sciences President’s International Fellowship Initiative (PIFI),
Grant No. 2021PM0066, the Chinese Ministry of Science and Technology
Foreign Expert Project, Grant No. QN20200143003, and the National
Natural Science Foundation of China (NSFC) under Grant No. 12047555.
We acknowledge financial support by the FEDER2020 funds, 
project Ref. A-FQM-406-UGR20 and from MCIN/AEI/10.13039/501100011033, 
Project Ref. PID2020-114767GB-I00. JM would like to thank 
the Department of Science and Technology (DST), Government of 
India, for financial support through Grant No. SR/WOS-A/PM-6/2019(G).
Figures~\ref{fig:threeleg},\ref{fig:mass}, and \ref{fig:exchange} 
are made using the open software JaxoDraw~\cite{Jaxodraw} 
distributed under the GNU General Public license. 

\appendix

\section{Second-order RGPEP}
\label{ap:SecondOrder}

\subsection{General formulas}

Our notation closely resembles the notation used in
Ref.~\cite{Glazek:2012qj}. Letters $a$, $b$, and $x$
denote configurations of particles, {\it i.e.} collections
of quantum numbers that label particle operators. Each 
configuration can contain arbitrary number and kinds of 
particles. $\cH_{t\,ab}$ denotes the coefficient that 
multiplies the term in $\cH_t$ in which particles in 
configuration $b$ are annihilated and particles in 
configuration $a$ are created. For arbitrary $a$ and 
$b$, the RGPEP Eq.~(\ref{eq:RGPEPeq1}) gives
\beq
\cH_{t\,ab}'
\es
- \left( P_a^- - P_b^- \right)^2 \cH_{It\,ab}
+ \sum_x \left( P_a^- + P_b^- - 2 P_x^- \right)
  \cH_{It\,ax} \cH_{It\,xb}
\ ,
\eeq
where $\sum_x$ denotes the sum over all possible configurations
$x$ and the sum or integration over all quantum numbers in each
individual configuration, $P_a^-$, $P_b^-$, and $P_x^-$ are sums
of front-form energies of all particles in configurations $a$,
$b$, and $x$, respectively, and $\cH_{It} = \cH_t - \cH_f$.
Factoring out the 1st-order form factors from vertices,
\beq
\cH_{t \, ab}
\es
f_{a.b, t} \, \cG_{t\, ab}
\ ,
\eeq
where
\beq
f_{a.b, t}
\es
e^{ - t \left( P_a^- - P_b^- \right)^2 }
\ ,
\eeq
we get differential equation for $\cG_{t \, ab}$:
\beq
\cG_{t\, ab}'
\es
  \sum_x \left( P_a^- + P_b^- - 2 P_x^- \right)
  \frac{f_{a.x, t} f_{x.b, t}}{f_{a.b, t}}
  \, \cG_{t \, ax} \, \cG_{t \, ab}
\ .
\label{eq:htprime}
\eeq
One obtains $f_{a.x, t} f_{x.b, t} / f_{a.b, t} =
\exp[ - 2 t (P_a^- - P_x^-)(P_b^- - P_x^-) ]$.
Expanding $\cG_t$ in a series in powers of the coupling
constant $g$,
\beq
\cG_{t \, ab}
\es
\cG_{f \, ab} + g \cG_{t1 \, ab} + g^2 \cG_{t2 \, ab} + \dots
\ ,
\eeq
and can solve Eq.~(\ref{eq:htprime}) order by order.
$\cG_f$ is independent of $t$, which is already reflected
in the notation. Up to second order,
\beq
\cG_{t1 \, ab} \es \cG_{01 \, ab} \ ,
\\
\cG_{t2 \, ab}
\es
  \cG_{02 \, ab}
+ \sum_x B_{t \, axb}^{(123,0)} \cG_{01 \, ax} \cG_{01 \, xb}
\ ,
\label{eq:RGPEPsol2nd}
\eeq
where
\beq
B_{t \, axb}^{(123,0)}
\es
\int_0^t d\tau
  \left( P_a^- + P_b^- - 2 P_x^- \right)
  \frac{f_{a.x, \tau} f_{x.b, \tau}}{f_{a.b, \tau}}
\ .
\eeq
After integration,
\beq
B_{t \, axb}^{(123,0)}
\es
\begin{cases}
\left( P_a^- + P_b^- - 2 P_x^- \right) t \ 
~~{\rm when}~~
\left( P_a^- - P_x^- \right) \left( P_b^- - P_x^- \right)= 0 \ ,
\\
\frac{1}{2}
\left[ (P_a^- - P_x^-)^{-1} + (P_b^- - P_x^-)^{-1} \right]
\left(
1 - f_{a.x, t} f_{x.b, t}/f_{a.b, t}
\right) ~~~~
\textrm{otherwise.}
\end{cases}
\eeq

\subsection{Self-interaction terms\label{sec:massDetail}}

For mass terms we have $P_a^- = P_b^-$, hence, $f_{a.b, t} = 1$,
and $f_{a.x, t} = f_{x.b, t}$.
Equation~(\ref{eq:RGPEPsol2nd}) becomes,
\beq
\cG_{t2 \, ab}
\es
  \cG_{02 \, ab}
+ \int_{\tilde i 3}
  \frac{1}{P_a^- - P_x^-}
  \left( 1 - f_{a.x, t}^2 \right)
  \cG_{01 \, ax} \cG_{01 \, xb}
\ ,
\label{eq:RGPEPsol2ndMass}
\eeq
where $\cG_{01 \, ax} = \bar u_i \gamma^\mu u_{\tilde i}
t^3_{i \tilde i} f_{i. \tilde i 3, t_r}
\tdelta_{i.\tilde i 3} \varepsilon_{3\mu}$,
$\cG_{01 \, xb} = \bar u_{\tilde i} \gamma^\nu u_{i'}
t^3_{\tilde i i'} f_{\tilde i 3.i', t_r}
\tdelta_{\tilde i 3.i'} \varepsilon_{3\nu}^*$,
and $P_a^- - P_x^- = (m_i^2 - \cM_{\tilde i 3}^2)/p_i^+$.
Therefore,
\beq
\cG_{t2 \, ab}
\es
  \cG_{02 \, ab}
+ \tdelta_{i.i'}
  \int_{\tilde i 3}
  t^3_{i \tilde i} t^3_{\tilde i i'}
  \frac{p_i^+ \tdelta_{\tilde i 3. i}}{m_i^2 - \cM_{\tilde i 3}^2}
  \left( f_{\tilde i 3.i, t_r}^2 - f_{\tilde i 3.i, t + t_r}^2 \right)
  \bar u_i \slashed\varepsilon_{3} u_{\tilde i}
  \bar u_{\tilde i} \slashed\varepsilon_{3}^* u_{i'}
\ .
\eeq
Now, using $\sum_{c_3} t^{c_3}_{i \tilde i} t^{c_3}_{\tilde i i'}
= C_F \delta_{c_i,c_{i'}}$, and $\sum_{\sigma_{\tilde i}}
\tdelta_{i.i'} \bar u_i \slashed\varepsilon_{3} u_{\tilde i}
\bar u_{\tilde i} \slashed\varepsilon_{3}^* u_{i'}
= \sum_{\sigma_{\tilde i}} \tdelta_{i.i'} \delta_{\sigma_i,\sigma_{i'}}
\ \bar u_i \slashed\varepsilon_{3} u_{\tilde i}
\bar u_{\tilde i} \slashed\varepsilon_{3}^* u_{i}$,
we get,
\beq
\cG_{t2 \, ab}
\es
  \cG_{02 \, ab}
+ \tdelta_{i.i'}
  \delta_{\sigma_i, \sigma_{i'}}
  \delta_{c_i, c_{i'}}
  \ C_F \sum_{\sigma_{\tilde i}, \sigma_3} \int[\tilde i 3]
  p_i^+ \tdelta_{\tilde i 3. i}
  \frac{f_{\tilde i 3.i, t + t_r}^2 - f_{\tilde i 3.i, t_r}^2}
  {\cM_{\tilde i 3}^2 - m_i^2}
  \bar u_i \slashed\varepsilon_{3} u_{\tilde i}
  \bar u_{\tilde i} \slashed\varepsilon_{3}^* u_i
\\
\es
  \cG_{02 \, ab}
+ \tdelta_{i.i'}
  \delta_{\sigma_i, \sigma_{i'}}
  \delta_{c_i, c_{i'}}
  \left[ I_i(t + t_r, m_g) - I_i(t_r, m_g) \right]
 .
\eeq
Writing $\cG_{02 \, ab} = \tdelta_{i.i'}
\delta_{\sigma_i, \sigma_{i'}} \delta_{c_i, c_{i'}} \delta m_{iX}^2$,
we obtain $\cG_{t2 \, ab} = \tdelta_{i.i'} \delta_{\sigma_i, \sigma_{i'}}
\delta_{c_i, c_{i'}} \delta m_{it}^2$, where $\delta m_{it}^2$
is defined in Eq.~(\ref{eq:effectiveMass}). Finally,
\beq
H_{\delta m}
\es
  \int_{11'} \cG_{t2 \, ab} \, b_{t1}^\dag b_{t1'}
+ \int_{22'} \cG_{t2 \, ab} \, d_{t2}^\dag d_{t2'}
\\
\es
  \int_1 \frac{\delta m_{1t}^2}{p_1^+} \, b_{t1}^\dag b_{t1}
+ \int_2 \frac{\delta m_{2t}^2}{p_2^+} \, d_{t2}^\dag d_{t2}
\ .
\eeq

\subsection{Gluon exchange terms\label{sec:exchDetail}}

For gluon exchange potentials, if gluon is emitted from
the quark, we have,
\beq
\cG_{01 \, ax}
\es
- j_2^\mu \ t^3_{2'2} \ f_{2, t_r}
  \ \tdelta_{2'3.2} \ \varepsilon_{3\mu}
\ ,
\\
\cG_{01 \, xb}
\es
  j_1^\nu \ t^3_{11'} \ f_{1, t_r}
  \ \tdelta_{13.1'} \ \varepsilon_{3\nu}^*
\ ,
\\
B_{t \, axb}^{(123,0)}
\es
\frac{1}{2}
\left(
  \frac{1}{p_2^- - p_{2'}^- - p_3^-}
+ \frac{1}{p_{1'}^- - p_1^- - p_3^-}
\right)
\left(
1 - \frac{f_{2.2'3, t} f_{13.1', t}}{f_{12.1'2', t}}
\right)
\\
\es
\frac{1}{2}
\left(
  \frac{p_3^+}{q_2^2 - m_g^2}
+ \frac{p_3^+}{q_1^2 - m_g^2}
\right)
\left(
1 - \frac{f_{2, t} f_{1, t}}{f_t}
\right)
\ ,
\eeq
and
\beq
\cG_{t2 \, ab}
\es
  \cG_{02 \, ab}
+ \int_3 \, B_{t \, axb}^{(123,0)}
  \, \cG_{01 \, ax} \, \cG_{01 \, xb}
\\
\es
  \cG_{02 \, ab}
- f_{1, t_r} f_{2, t_r}
  \tdelta_{12.1'2'}
\frac{1}{2}
\left(
  \frac{1}{q_1^2 - m_g^2}
+ \frac{1}{q_2^2 - m_g^2}
\right)
\left(
1 - \frac{f_{1, t} f_{2, t}}{f_t}
\right)
  \sum_{\sigma_3} \varepsilon_{3\mu} \varepsilon_{3\nu}^*
  \ j_1^\nu \, j_2^\mu
  \ t^a_{11'} t^a_{2'2}
\ . 
\nn
\label{eq:cGt2abexch}
\eeq
The initial condition, $\cG_{02 \, ab}$, includes the canonical
instantaneous interaction (regularized) plus the counterterm,
\beq
\cG_{02 \, ab}
\es
\tdelta_{12.1'2'}
\left(
- f_{1, t_r} f_{2, t_r} \, \frac{ j_1^+ j_2^+ }{ (q^+)^2 }
+ X
\right)
t^a_{11'} t^a_{2'2}
\ .
\eeq
If gluon is emitted from the antiquark, we have,
\beq
\cG_{01 \, ax}
\es
  j_1^\mu \ t^3_{11'} \ f_{1, t_r}
  \ \tdelta_{1'3.1} \ \varepsilon_{3\mu}
\ ,
\\
\cG_{01 \, xb}
\es
- j_2^\nu \ t^3_{2'2} \ f_{2, t_r}
  \ \tdelta_{23.2'} \ \varepsilon_{3\nu}^*
\ ,
\\
B_{t \, axb}^{(123,0)}
\es
\frac{1}{2}
\left(
  \frac{1}{p_1^- - p_{1'}^- - p_3^-}
+ \frac{1}{p_{2'}^- - p_2^- - p_3^-}
\right)
\left(
1 - \frac{f_{1.1'3, t} f_{23.2', t}}{f_{12.1'2', t}}
\right)
\\
\es
\frac{1}{2}
\left(
  \frac{p_3^+}{q_1^2 - m_g^2}
+ \frac{p_3^+}{q_2^2 - m_g^2}
\right)
\left(
1 - \frac{f_{1, t} f_{2, t}}{f_t}
\right)
\ ,
\eeq
and and we arrive again at Eq.~(\ref{eq:cGt2abexch}). 
The final result,
\beq
H_{Ut}
\es
\int_{121'2'} \cG_{t2 \, ab}
\ b_{t\,1}^\dag \, d_{t\,2}^\dag \, d_{t\,2'} \, b_{t\,1'}
\ ,
\eeq
after simple manipulations gives Eq.~(\ref{eq:HUt}).

\section{Gluon exchange counterterm}
\label{sec:gluonExchCounterterm}

If we split $H_{Ut} = \hat U_C + \hat U_H + \hat U_X$,
in accordance with Eq.~(\ref{eq:kernelUt}), then
\beq
\bra{L} \hat U_X \ket{R}
\es
- C_F g_t^2
  \tdelta_{P_L.P_R}
  \sum_{\sigma_1, \sigma_2}
  \int\frac{dx_1 d^2 k_{12}^\perp}{16\pi^3 x_1 x_2}
  \, \psi_L^*(1,2)
  \int\frac{dx_{1'} d^2 k_{1'2'}^\perp}{16\pi^3}
  \, \left( Y_1 + Y_2 + Y_3 \right)
\ ,
\eeq
where 
\beq
Y_1
\es
  f_{1, t_r} f_{2, t_r}
  \, [ Z(x_{1'}) - Z(x_{1})] /q^{+2} 
\ ,
\\
Y_2
\es
  \left[ f_{1, t_r} f_{2, t_r}
        - e^{-2 t_r \frac{(\Delta k^2 + m_g^2)^2}{P^{+2} x_3^2}} \right]
  \, Z(x_{1}) / q^{+2}
\\
Y_3
\es
  e^{-2 t_r \frac{(\Delta k^2 + m_g^2)^2}{P^{+2} x_3^2}} 
  \, Z(x_{1})/ q^{+2}
- \frac{f_t \, X}{x_{1'} x_{2'}}
  \, \psi_R(x_{1'}, k_{1'2'})
\ ,
\eeq
and
\beq
Z(x_{1'})
\es
  \left( 1 + \frac{ q_1^2 + q_2^2 }{ 2 } \, \cF \right)
  \frac{f_t \, j_{1}^+ j_{2}^+}{x_{1'} x_{2'}}
  \, \psi_R(x_{1'}, k_{1'2'})
\ ,
\\
Z(x_1)
\es
\lim_{x_{1'} \to x_1} Z(x_{1'})
\rs
  4 P^{+2} \tilde f_t
  \, \frac{ m_g^2 }{ m_g^2 + \Delta k^2 }
  \, \psi_R(x_{1}, k_{1'2'})
\ ,
\eeq
with the dependence of $Z$ on $x_{1}$, $k_{12}$, and $k_{1'2'}$
not indicated explicitly, while $\tilde f_t = \lim_{x_{1'} \to x_1} f_t$.

Lifting the regularization we obtain
\beq
\lim_{t_r \to 0^+}
\int\frac{dx_{1'} d^2 k_{1'2'}^\perp}{16\pi^3} \, Y_1
\es
\cP\int dx_{1'} \int\frac{d^2 k_{1'2'}^\perp}{16\pi^3}
\lim_{t_r \to 0^+} Y_1
\ .
\eeq
The singularity $1/(q^+)^2$ is removed due to the difference 
$Z(x_{1'}) - Z(x_{1})$. The principal value $\cP$ is obtained 
because the regulator is approximately symmetric in $q^+$ in
the vicinity of $q^+ = 0$. We take the limit $m_g \to 0$,
\beq
\lim_{m_g \to 0^+}
\cP\int dx_{1'} \int\frac{d^2 k_{1'2'}^\perp}{16\pi^3}
\lim_{t_r \to 0^+} Y_1
\es
\cP\int dx_{1'} \int\frac{d^2 k_{1'2'}^\perp}{16\pi^3}
\lim_{m_g \to 0^+}
[ Z(x_{1'}) - Z(x_{1})]/ q^{+2}
\\
\es
- \cP\int dx_{1'} \int\frac{d^2 k_{1'2'}^\perp}{16\pi^3}
  \frac{ 1 }{ q^{+2} }
  \frac{(q_1^2 - q_2^2)^2}{4 q_1^2 q_2^2}
  \frac{f_t \, j_{1}^+ j_{2}^+}{x_{1'} x_{2'}}
  \, \psi_R(x_{1'}, k_{1'2'})
\nn
\\
\es
- \int dx_{1'} \int\frac{d^2 k_{1'2'}^\perp}{16\pi^3}
  \frac{(\cM_{12}^2 - \cM_{1'2'}^2)^2}{4 q_1^2 q_2^2}
  \frac{f_t \, j_{1}^+ j_{2}^+}{p_{1'}^+ p_{2'}^+}
  \, \psi_R(x_{1'}, k_{1'2'})
\ ,
\nn
\eeq
where in the last equality the principal value turns out not 
needed anymore. The term $Y_2$ gives zero in the limit 
$t_r \to 0$. The term $Y_3$ is where the divergence
resides and needs a counter term. Below we present
the demonstration that small-$x$ divergences are canceled 
once the counter term is added, which we denote by $X$. 
Using Eq.~(\ref{eq:exchCT}) and the definition of $Z$ we rewrite $Y_3$,
\beq
Y_3
\es
  e^{-2 t_r \frac{(\Delta k^2 + m_g^2)^2}{P^{+2} x_3^2}} 
  \, Z(x_{1}) / q^{+2}
- \frac{1}{P^+} \delta(x_{1'} - x_{1})
  \, \frac{1}{\Delta k^2 + m_g^2}
  \, \sqrt{\frac{\pi}{2 t_r}}
  \, Z(x_1)
\ .
\eeq
The integral becomes,
\beq
\int\frac{dx_{1'} d^2 k_{1'2'}^\perp}{16\pi^3} \, Y_3
\es
\int\frac{d^2 k_{1'2'}^\perp}{16\pi^3}
\, \frac{ Z(x_{1}) }{ P^{+2} }
\int_0^1 dx_{1'}
\left[
  \frac{ e^{-\frac{2 t_r (\Delta k^2 + m_g^2)^2}{P^{+2} (x_{1'} - x_1)^2}} }{ (x_{1'} - x_1)^2 }
- \delta(x_{1'} - x_{1})
  \, \frac{P^+}{\Delta k^2 + m_g^2}
  \, \sqrt{\frac{\pi}{2 t_r}}
\right] .
\nn
\eeq
The integral over $x_{1'}$ can be evaluated,
\beq
\int\frac{dx_{1'} d^2 k_{1'2'}^\perp}{16\pi^3} \, Y_3
\es
\int\frac{d^2 k_{1'2'}^\perp}{16\pi^3}
\, \frac{ Z(x_{1}) }{ P^{+2} }
\, \frac{P^+}{\Delta k^2 + m_g^2}
\, \sqrt{\frac{\pi}{2 t_r}}
\nt
\left[
  \frac{1}{2}
  \mathrm{erfc}\left(\frac{\sqrt{2 t_r} (\Delta k^2 + m_g^2)}{P^+ x_2}\right)
+ \frac{1}{2}
  \mathrm{erfc}\left(\frac{\sqrt{2 t_r} (\Delta k^2 + m_g^2)}{P^+ x_1}\right)
- 1
\right] . \qquad
\eeq
Note that $-1$ in the square bracket comes from the
counter term. If it were absent, the bracket would be $1$ for
$t_r \to 0$ and the integral would be proportional to
$t_r^{-1/2}$. Therefore, the matrix elements of $\hat U_X$
are divergent without the counter term.
Expansion of the square bracket for small $t_r$ gives,
\beq
\int\frac{d^2 k_{1'2'}^\perp}{16\pi^3}
\, \frac{ Z(x_{1}) }{ P^{+2} }
\, \frac{P^+}{\Delta k^2 + m_g^2}
\, \sqrt{\frac{\pi}{2 t_r}}
\left[
- \sqrt{\frac{2 t_r}{\pi}}
  \left(
    \frac{\Delta k^2 + m_g^2}{P^+ x_2}
  + \frac{\Delta k^2 + m_g^2}{P^+ x_1}
  \right)
+ O\left(t_r^{3/2}\right)
\right] .
\eeq
Hence,
\beq
\lim_{t_r \to 0^+}
\int\frac{dx_{1'} d^2 k_{1'2'}^\perp}{16\pi^3} \, Y_3
\es
- \int\frac{d^2 k_{1'2'}^\perp}{16\pi^3}
  \, \frac{ Z(x_{1}) }{p_1^+ p_2^+}
\ . \qquad
\eeq
Since $Z(x_1) \to 0$ when $m_g \to 0$,
\beq
\lim_{m_g \to 0^+}
\lim_{t_r \to 0^+}
\int\frac{dx_{1'} d^2 k_{1'2'}^\perp}{16\pi^3} \, Y_3
\es
0
\ .
\eeq
Summarizing,
\beq
\lim_{m_g \to 0^+}
\lim_{t_r \to 0^+}
\bra{L} \hat U_X \ket{R}
\es
  C_F g_t^2
  \tdelta_{P_L.P_R}
  \sum_{\sigma_1, \sigma_2}
  \sum_{\sigma_{1'}, \sigma_{2'}}
  \int\frac{dx_1 d^2 k_{12}^\perp}{16\pi^3 x_1 x_2}
  \int\frac{dx_{1'} d^2 k_{1'2'}^\perp}{16\pi^3 x_{1'} x_{2'}}
\nt
  \psi_L^*(1,2)
  \frac{(\cM_{12}^2 - \cM_{1'2'}^2)^2}{4 q_1^2 q_2^2}
  \frac{f_t \, j_{1}^+ j_{2}^+}{P^{+2}}
  \, \psi_R(1',2')
\ .
\eeq
This formula could be obtained by first taking the limit
$t_r \to 0$ in $U_X$, neglecting the counter term and
the divergences, and then taking the limit $m_g \to 0$.
Such procedure is valid only if the vicinity of $q^+ = 0$
is excluded. However, the above analysis shows that if one
includes the counter term the result is the same. Thus, 
we can write
\beq
\lim_{m_g \to 0^+}
\lim_{t_r \to 0^+}
U_X
\es
- f_t \, \sqrt{x_1 x_2 x_{1'} x_{2'}}
  \, \frac{(\cM_{12}^2 - \cM_{1'2'}^2)^2}{q_1^2 q_2^2}
  \, \delta_{\sigma_1, \sigma_{1'}}
  \, \delta_{\sigma_2, \sigma_{2'}}
\ ,
\eeq
and in the nonrelativistic limit,
\beq
V_X
\es
\lim_\textrm{NR}
\lim_{m_g \to 0^+}
\lim_{t_r \to 0^+}
\frac{U_X}{4 m_1 m_2}
\\
\es
- f_t
  \frac{(m_1 + m_2)^2}{4 (m_1 m_2)^2}
  \frac{\left( \vec k^2 - \vec k'^2 \right)^2}{(\vec q\,^2)^2}
  \, \delta_{\sigma_1, \sigma_{1'}}
  \, \delta_{\sigma_2, \sigma_{2'}}
\rs
  O\left( \frac{1}{\mu^2} \right) .
\label{eq:VXNR}
\eeq

\bibliography{references}

\end{document}